\documentclass[aps,prd,showpacs,twocolumn,amsmath,10pt,superscriptaddress,floatfix,nofootinbib]{revtex4-1}

\usepackage{epsfig,amssymb,amsfonts,bm,color}
\usepackage{epstopdf}

\graphicspath{{figs.dir/}}

\newcommand{\be}{\begin{equation}}
\newcommand{\ee}{\end{equation}}
\newcommand{\bea}{\begin{eqnarray}}
\newcommand{\eea}{\end{eqnarray}}
\newcommand{\beas}{\begin{eqnarray*}}
\newcommand{\eeas}{\end{eqnarray*}}
\newcommand{\ds}{\displaystyle}
\newcommand{\vep}{{\bm p}}
\newcommand{\vek}{{\bm k}}
\newcommand{\veq}{{\bm q}}
\newcommand{\tv}{t^v}

\newcommand{\tw}{t^w}
\newcommand{\VI}{V_0}
\newcommand{\VII}{V_1}
\newcommand{\dq}{d^3q}
\newcommand{\Br}{\mbox{Br}}
\newcommand{\calg}{{\cal G}}
\newcommand{\Sa}{{\cal S}_a}
\newcommand{\g}{S}
\newcommand{\Ne}{N_\text{e}}
\newcommand{\Nin}{N_\text{in}}
\newcommand{\Np}{N_\text{p}}
\newcommand{\Nv}{N_\text{v}}

\DeclareMathSymbol{\varGamma}{\mathord}{letters}{"00}

\begin{document}

\title{Interplay of quark and meson degrees of freedom in near-threshold states: \\
A practical parametrisation for line shapes}

\author{F.-K. Guo}
\affiliation{State Key Laboratory of Theoretical Physics, Institute of Theoretical Physics, Chinese Academy of Sciences,
Beijing 100190, China}
\affiliation{Helmholtz-Institut f\"ur Strahlen- und Kernphysik and Bethe
Center for Theoretical Physics, \\Universit\"at Bonn,  D-53115 Bonn, Germany}

\author{C. Hanhart}
\affiliation{Forschungszentrum J\"ulich, Institute for Advanced Simulation, Institut f\"ur Kernphysik (Theorie) and
J\"ulich Center for Hadron Physics, D-52425 J\"ulich, Germany}

\author{Yu.S. Kalashnikova}
\affiliation{Institute for Theoretical and Experimental Physics, 117218,
B.Cheremushkinskaya 25, Moscow, Russia}

\author{P. Matuschek}
\affiliation{Forschungszentrum J\"ulich, Institute for Advanced Simulation, Institut f\"ur Kernphysik (Theorie) and
J\"ulich Center for Hadron Physics, D-52425 J\"ulich, Germany}

\author{R.V. Mizuk}
\affiliation{Lebedev Physics Institute, 119991, Leninsky prospect 53, Moscow, Russia}
\affiliation{National Research Nuclear University MEPhI, 115409, Kashirskoe highway 31, Moscow, Russia}
\affiliation{Moscow Institute of Physics and Technology, 141700, 9 Institutsky lane, Dolgoprudny, Moscow Region, Russia}

\author{A.V. Nefediev}
\affiliation{Institute for Theoretical and Experimental Physics, 117218,
B.Cheremushkinskaya 25, Moscow, Russia}
\affiliation{National Research Nuclear University MEPhI, 115409, Kashirskoe highway 31, Moscow, Russia}
\affiliation{Moscow Institute of Physics and Technology, 141700, 9 Institutsky lane, Dolgoprudny, Moscow Region, Russia}

\author{Q. Wang}
\affiliation{Helmholtz-Institut f\"ur Strahlen- und Kernphysik and Bethe
Center for Theoretical Physics, \\Universit\"at Bonn,  D-53115 Bonn, Germany}
\affiliation{Forschungszentrum J\"ulich, Institute for Advanced Simulation, Institut f\"ur Kernphysik (Theorie) and
J\"ulich Center for Hadron Physics, D-52425 J\"ulich, Germany}

\author{J.-L. Wynen}
\affiliation{Forschungszentrum J\"ulich, Institute for Advanced Simulation, Institut f\"ur Kernphysik (Theorie) and
J\"ulich Center for Hadron Physics, D-52425 J\"ulich, Germany}

\begin{abstract}
We propose a practical parametrisation for the line shapes of near-threshold states compatible with all
requirements of unitarity and analyticity. The coupled-channel system underlying the proposed parametrisation includes
bare poles and an arbitrary number of elastic and inelastic channels treated fully nonperturbatively. The resulting
formulae are general enough to be used for a simultaneous analysis of the data in all available production and decay
channels of the (system of) state(s) under consideration for a quite wide class of reactions. As an example, we fit
the experimental data currently available for several decay channels for the charged $Z_b^{(\prime)}$
states in the spectrum of bottomonia and find a good overall description of the data.
We find the present data to be consistent with the $Z_b(10610)$ as a virtual state and with the $Z_b(10650)$ as a
resonance, both residing very close to the $B\bar{B}^*$ and $B^*\bar{B}^*$ threshold, respectively.
\end{abstract}


\maketitle

\section{Introduction}

In the last decades an enormous bulk of data on the charmonia((like)) and bottomonia((like)) states lying above the
open-flavour thresholds have been collected by many experiments, such as
BABAR, Belle, BESIII, CDF, D\O, and LHCb. Future
high luminosity experiments, in particular, the forthcoming experiment Belle-II at KEK and PANDA at FAIR,
are expected to provide new high
precision and high statistics data for already known states, as well as for new, yet unobserved ones
in various final states
\cite{Brambilla:2010cs,Brambilla:2014jmp,Abe:2010gxa,Drutskoy:2012gt,Asner:2008nq,Lutz:2009ff}.
Traditionally data in different channels were analysed individually by use of Breit-Wigner distributions, or
sums thereof, combined with some background function. However, this procedure provides only limited
information on the state studied: first of all, Breit-Wigner parameters are reaction-dependent; second,
summing Breit-Wigners in general violates unitarity, and last but not least, by studying individual channels
only, one does not exploit the full information content provided by the measurements.
In particular, the theoretical description of the states above the open-flavour thresholds calls
for using adequate parametrisations for the line shapes which should be capable of describing such phenomena as finite
widths of the
constituents, multiple thresholds in the vicinity of the resonances, an interplay of the quark and hadron degrees of
freedom in near-threshold states, and so on. In the meantime, such parametrisations need to be easy
to handle in order to be useful for the analysis of experimental data.

Consider first an unstable particle coupled to the had\-ro\-nic channel, open at $E=0$, with the
coupling constant $g_f$. In the effective range approximation the scattering amplitude can be written in the
form \cite{Flatte:1976xu}
\be
{\cal M}(E)=\frac{g_f/2}{E-E_0+i (g_f/2)k},
\label{FE1}
\ee
where the momentum $k$ is
$$
k(E)=\sqrt{2\mu E}\Theta(E)+i\sqrt{-2\mu E}\Theta(-E),
$$
and $\mu$ is the reduced mass in the hadronic channel. Equation (\ref{FE1}) can be viewed as the Breit-Wig\-ner
amplitude with the momentum dependence of the elastic width taken into account explicitly and it is valid,
if the nearest additional threshold, located at $E=\Delta$, is far away from the considered threshold at $E=0$, that is
$|\Delta|\gg |E_0|$. Also, the direct interaction in the hadronic channel should not generate additional near-threshold
poles in
the $S$-matrix. As soon as one of these conditions fails Eq.~(\ref{FE1}) has to be generalised.
In particular, in Ref.~\cite{Baru:2010ww} such a generalisation is given for the case when the direct interaction in
the hadronic channel does generate near-threshold poles in the $S$-matrix and a nontrivial interplay between quark and
meson degrees of freedom takes place. The resulting line shapes may have quite a peculiar form, drastically different
from the ones given by the simple Flatt{\'e} formula of Eq.~(\ref{FE1}).

Straightforward generalisations of Eq.~(\ref{FE1}) to the multichannel case are
discussed in
Refs.~\cite{Hanhart:2007yq,Kalashnikova:2009gt,Kalashnikova:2010zz,Braaten:2007dw,Stapleton:2009ey,Zhang:2009bv},
where all effects of the direct interaction between mesonic channels are
absorbed into the effective coupling constants. Mo\-re sophisticated
approaches to the direct interactions in the mesonic channels are employed in
Refs.~\cite{Artoisenet:2010va,Hanhart:2011jz}. Effects of the finite width of the
constituents are discussed in Refs.~\cite{Kalashnikova:2009gt,Braaten:2007dw,Hanhart:2010wh}.
{Related discussions can also be found in Ref.~\cite{Meng:2014ota}.}

In this paper we further extend the basis of states involved and consider a
coupled-channel problem for near-threshold phenomena in a physical system which
contains not only near-threshold poles and allows for additional elastic (in
the example below, nearby open-flavour) meso\-nic channels, but also
incorporates inelastic (in the example below, more distant hidden-flavour)
channels fully nonperturbatively as required by unitarity. The resulting system
of equations is expected to be rich enough to provide a realistic description of
the line shapes for a quite wide class of reactions.

The formalism used is set up in a very general way. In particular, we allow
for the inclusion of a set of bare poles in addition to various
nonperturbative rescatterings. Effectively this provides an additional momentum- and
ener\-gy-dependent interaction and therefore gives an additional flexibility for the fitting of experimental data,
but does not \emph{a priori} impose
any assumption on the nature/wave function decomposition of a given state.
In particular, with the pole terms included it becomes easily possible to also
analyse systems with resonances above the thresholds. The main ideas
as well as the key results have already been presented in
Ref.~\cite{Hanhart:2015cua}---here much more detailed derivations and discussions
are presented and the updated experimental data are analysed. In addition, we
briefly discuss the possible role of the one-pion exchange.

For illustration of the formalism below we study decays of a system that
contains a $\bar{Q}Q$ pair, with $Q$ denoting a heavy quark. We refer to the
open-flavour channels $(\bar{q}Q)(\bar{Q}q)$ (here $q$ denotes a light quark) by
greek letters $\alpha$, $\beta$, $\ldots$ and to the hidden-flavour channels
$(\bar{Q}Q)(\bar{q}q)$ by latin letters $i$, $j$, $\ldots$.
The explicit poles are included as additional channels labelled by latin letters from the beginning of the alphabet,
that is $a$, $b$, $\ldots$.

Paradigmatic examples of such physical systems are, e.g., the $X(3872)$ decaying into
the open-charm channels $D\bar{D}^*$ \cite{Gokhroo:2006bt} and the hidden-charm channels $\pi^+\pi^-
J/\psi$
\cite{Choi:2003ue} and $\pi^+\pi^-\pi^0 J/\psi$ \cite{delAmoSanchez:2010jr}, or
 $Z_b(10610)$ and $Z_b(10650)$  decaying into the
$B^{(*)}\bar{B}^*$ open-bottom \cite{Adachi:2012cx} and $\pi\Upsilon(nS)/\pi
h_b(mP)$ ($n=1,2,3$, $m=1,2$) hidden-bottom \cite{Belle:2011aa} channels.
While additional effects such as finite widths of the constituents and
additional interactions between outgoing particles may also play a role and thus may have
to be included on top of the interactions considered in this work (for a recent
discussion of such effects see Ref.~\cite{Chen:2015jgl}), nevertheless the gross
features of the coupled-channel problem are captured by the presented formalism and the parametrisation based on it is expected to be realistic.

\section{Solution of the Lippmann-Schwinger equation}\label{solLS}

\subsection{Simplification of the Lippmann-Schwinger equation in a two-channel toy model}
\label{sec:toy}

For the case of the structures near the open-flavour thresholds, as we will
show, the coupled-channel Lippmann-Schwinger equation (LSE) can be simplified
by absorbing some channels into the definition of an effective potential. To see
how this works, it is instructive to study a simple two-channel toy model.
In this subsection, we write the LSE in the operator form for simplicity. It
will be written more explicitly in the form of integral equations in the next subsection.

Let us start with the LSE for the $t$ matrix
\be
t = v-vSt,
\ee
where $S$ is the matrix for the free Green's functions in the channel space.
The potential is parametrised as
\be
v = \begin{pmatrix}
v_{11} & v_{12} \\
v_{21} & v_{22}
\end{pmatrix}.
\ee
Note that time reversal invariance demands that $v_{12}=v_{21}$ while
for $t$ to be unitary, all $v_{ij}$'s must be real.
Explicitly, we have a system of four coupled-channel equations
\be
t_{ij}=v_{ij}-\sum_{k=1,2}v_{ik}S_kt_{kj},\quad i,j=1,2,
\ee
which, however, effectively reduce to single-channel equations if any of the potentials $v_{ij}$ vanishes. The two channels decouple from each
other trivially if the off-diagonal components are set to zero, $v_{12}=v_{21}=0$. Let us now focus on the case of a vanishing diagonal matrix
element of $v$. For definiteness, we set $v_{22}=0$. Then the $t$ matrix components $t_{12}$, $t_{21}$, and $t_{22}$ can be expressed through
the component $t_{11}$ straightforwardly,
\begin{eqnarray}
t_{12}&=&v_{12}-t_{11}S_1v_{12},\nonumber\\
t_{21}&=&v_{21}-v_{21}S_1t_{11},\label{eq:toy}\\
t_{22}&=&-v_{21}S_1v_{12}+v_{21}S_1t_{11}S_1v_{12},\nonumber
\end{eqnarray}
while $t_{11}$ comes as a solution of a single-channel LSE
\be
t_{11}=V_{11}-t_{11}S_1V_{11}=V_{11}-V_{11}S_1t_{11},
\ee
with the effective potential
\be
V_{11} = v_{11}-v_{12}S_2v_{21},
\label{V11eff}
\ee
which admits a transparent physical interpretation: elastic scattering in channel 1 proceeds either through the direct interaction potential $v_{11}$
or due to the transition through channel 2.

One sees therefore that channel 2 only enters additively in the effective potential, generalisation to additional channels being trivial.
This simplification can be applied to the case studied here because the interaction
between a light hadron and a heavy quarkonium is OZI forbidden and therefore it is very
weak. We discuss a realistic case in the following sections.

\subsection{Solution of the multichannel Lippmann-Schwinger equation}

In this subsection we formulate a multichannel model and solve the corresponding Lippmann-Schwinger equations using the simplifying trick described
in the previous subsection.

The key ingredients of the model are (i) the direct interaction in the
open-flavour channels described by the potential $v_{\alpha\beta}(\vep,\vep')$
as well as that in the hidden-flavour channels $v_{ij}(\vek,\vek')$, (ii) the transition form factor between the
open-flavour and hidden-flavour
channels\footnote{A microscopic model for this interaction can be found, for example, in
Refs.~\cite{Danilkin:2011sh,Danilkin:2009hr}.}
\be
v_{\alpha i}(\vep,\vek),\quad \alpha=\overline{1,\Ne},\quad i=\overline{1,\Nin},
\label{v}
\ee
and, finally,
(iii) the transition form factors between the bare pole terms and the
open-flavour and hidden-flavour channels,
\be
v_{a\alpha}(\vep)\quad \text{and}\quad v_{ai}(\vek), \quad a = \overline{1,\Np},
\label{vpole}
\ee
respectively.
The open-flavour and hidden-flavour channels will alternatively be called elastic
and inelastic channels, respectively. Note that unitarity in combination with the
T-invariance calls for a real and symmetric scattering potential, as long as all relevant channels are included
explicitly in the
model. Actually, one can reverse this statement: if a high-quality fit for the data
demands that some of the parameters be complex, the model should be regarded as incomplete.
Thus, the formalism outlined here implicitly provides a diagnostic tool to investigate,
whether or not for certain states all relevant channels are already discovered/included.

The interaction potential can be summarised in the form
\be
\raisebox{-3mm}{$\hat{V}=$}
\begin{tabular}{cc}
\scriptsize $b = \overline{1,\Np}\hspace*{5mm}
\beta=\overline{1,\Ne}\hspace*{7mm}i=\overline{1,\Nin}$&\\[2mm] $
\begin{pmatrix}
v_{ab}   & v_{a\beta}(\vep')& v_{ai}(\vek)\\[2mm]
v_{\alpha b}(\vep)~&~v_{\alpha\beta}(\vep,\vep')~&~v_{\alpha
i}(\vep,\vek)\\[2mm] v_{jb}(\vek')&v_{j\beta}(\vek',\vep')&v_{ji}(\vek',\vek)
\end{pmatrix}
$&
\begin{tabular}{c}
\scriptsize $a=\overline{1,\Np}$\\[2mm]
\scriptsize$\alpha=\overline{1,\Ne}$\\[2mm]
\scriptsize$j=\overline{1,\Nin}$.
\end{tabular}
\end{tabular}
\label{Vpot}
\ee
To simplify the notation we use the same symbol for incoming and outgoing
vertex functions---however, the context will always make it clear which one is
meant in a given equation.
The number of the elastic channels $\Ne$ and the number of the inelastic
channels $\Nin$ remain unspecified and can be chosen as large as suggested by
the particular reaction being analysed. For generality, we do not
specify the number of bare poles either. The coupled channel problem with
interaction potential (\ref{Vpot}) can be solved analytically, if a separable form of
the transition form factors $v_{\alpha i}(\vep,\vek)$ is assumed. However such a
general solution appears to be bulky and practically useless for the data
analysis, since it requires multiple inversions of large matrices of the
dimension $(\Ne+\Nin+\Np)\times(\Ne+\Nin+\Np)$.
Besides, inclusion of an additional inelastic channel requires the entire
procedure to be started from scratch.
Meanwhile, there are good reasons to neglect the direct interactions in the
inelastic channels. For example, for the systems we focus on here such interactions are expected to
be very weak---since there are no light quarks in heavy quarkonia, the
interaction of pions with them is OZI suppressed and it thus
vanishes at leading order in a low-energy
expansion for the pion-quarkonium interaction. This conjecture is confirmed by
the small values of the $\pi$-$\bar{Q}Q$
scattering lengths, which are estimated to be $\lesssim0.02$~fm~\cite{Liu:2012dv}
and found to be consistent with 0 in lattice QCD studies~\cite{Liu:2008rza,Detmold:2012pi}.
Therefore, in the following we set
$v_{ij}(\vek,\vek')\equiv 0$. As a result, the interaction potential of
Eq.~(\ref{Vpot}) reads
\be
\raisebox{-3mm}{$\hat{V}=$}
\begin{tabular}{cc}
\scriptsize
$\hspace*{-3mm}B=\overline{1,\,\Ne+\Np}\hspace*{5mm}i=\overline{1,\Nin}$&\\[2mm] $
\begin{pmatrix}
v_{AB}(\vep,\vep')~&~v_{Ai}(\vep,\vek)\\[2mm]
v_{jB}(\vek',\vep')&0
\end{pmatrix}
$&
\begin{tabular}{c}
\scriptsize$A=\overline{1\,\Ne+\Np}$\\[2mm]
\scriptsize$j=\overline{1,\Nin}$,
\end{tabular}
\end{tabular}
\label{Vpot2}
\ee
where, for convenience, we formally treat the pole terms as additional
elastic channels and use capital greek letters for the corresponding indices,
which now take values from 1 to $\Ne+\Np$.

The toy model from the previous subsection tells us that omission of
rescatterings within the inelastic channels introduces a great simplification
since they enter only additively in the effective potential.
Besides that we can completely disentangle elastic channels (including the pole
terms) and inelastic channels.
Consequently, solving the coupled-channel Lippmann-Schwinger equation amounts
to the inversion of matrices as small as $(\Ne+\Np)\times (\Ne+\Np)$ independent
of the number of inelastic channels---see Eq.~(\ref{tabnew1}). Furthermore, the formulae
to be derived below allow one to disentangle
the elastic channels from the bare poles too---see Eqs.~(\ref{ttt}), (\ref{tsol}),
(\ref{ta0sol})---so that eventually the problem reduces to inverting matrices as small as only $\Ne\times\Ne$ and
$\Np\times\Np$ independently. For $\Ne$ and $\Np$ smaller or equal to two, as in the case below, this
can be done straightforwardly in the explicit form.
Therefore the suggested approach guarantees a crucial simplification of the
calculations. In particular, it speeds up the codes drastically, making  combined analyses of
experimental data in various channels significantly easier.
Especially, adding an extra inelastic channel (explicitly or implicitly, through
an additional constant inelasticity) changes the final expressions only
marginally.

In order to formulate and solve the Lippmann-Schwin\-ger equation for the scattering $t$ matrix let us introduce the
effective interaction potential in the elastic channels [\emph{cf.} Eq.~(\ref{V11eff})]
\bea
&&V_{AB}(\vep,\vep')=v_{AB}(\vep,\vep')\nonumber\\
&&\hspace*{10mm}-\sum_i\int v_{A i}(\vep,\veq)\g_i(\veq)v_{iB}(\veq,\vep')\dq,
\label{Veff0}
\eea
where the quantity $\g_i(\veq)$ denotes the propagator of the $i$th $(\bar{Q}Q)(\bar{q}q)$ pair. The physical interpretation of
this
potential is straightforward: a transition from elastic channel $A$ to elastic channel $B$ proceeds either
through the direct interaction potential $v_{AB}(\vep,\vep')$ (including the pole terms) or through the inelastic
channels, where the sum in $i$
runs over all inelastic ``bubbles.'' Notice that Eq.~(\ref{Veff0}) as well as similar formulae below which
contain capital greek subscripts should be treated as schematic since, depending on a particular component of the
corresponding potential or of the $t$ matrix, the number of the arguments can be different---see
Eq.~(\ref{Vpot}). When written in components, potential (\ref{Veff0})
takes the form
\bea
V_{ab}&=& \raisebox{-0.8mm}{\epsfig{file=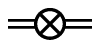,width=8mm}} -\sum_i
\raisebox{-2.7mm}{\epsfig{file=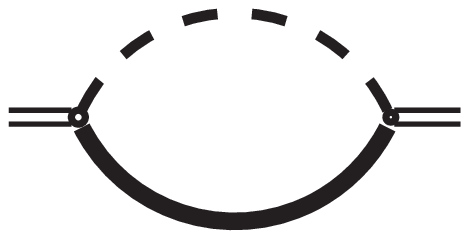,width=16mm}}\label{V00def}\\
&=& v_{ab} -\sum_i \int v_{ai}(\veq)\g_i(\veq)v_{ib}(\veq) \dq \nonumber\\
&\equiv&-{\cal G}_{0,ab}^\text{in},\nonumber
\eea
\bea
V_{\alpha
a}(\vep)&=&\raisebox{-3mm}{\epsfig{file=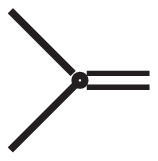,width=9mm}}
-\sum_i\raisebox{-2.7mm}{\epsfig{file=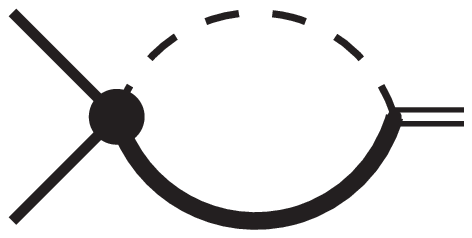,width=16mm}}\label{Va0}\\
&=&v_{\alpha a}(\vep)-\sum_i\int v_{\alpha
i}(\vep,\veq)\g_i(\veq) v_{i a}(\veq)\dq,\nonumber
\eea
\bea
V_{a\beta}(\vep)&=&\raisebox{-3mm}{\epsfig{file=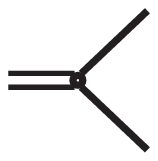,width=9mm}}
-\sum_i\raisebox{-2.7mm}{\epsfig{file=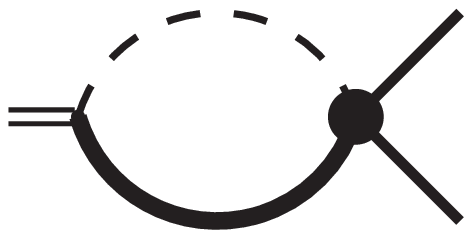,width=16mm}}\label{V0a}\\
&=&v_{a\beta}(\vep)-\sum_i\int
v_{ai}(\veq)\g_i(\veq)v_{i\beta}(\veq,\vep)\dq,\nonumber
\eea
\bea
&&\hspace*{-10mm}V_{\alpha\beta}(\vep,\vep')=\raisebox{-1.6mm}{\epsfig{file=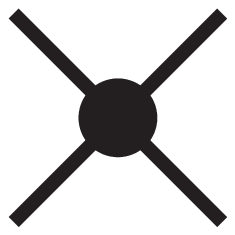,width=5.5mm}}-\sum_i
\raisebox{-3mm}{\epsfig{file=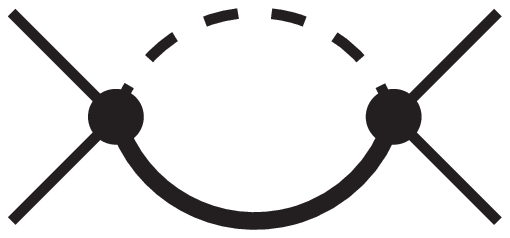,width=17mm}}\label{Vab}\\
&=&v_{\alpha\beta}(\vep,\vep')-\sum_i\int v_{\alpha i}(\vep,\veq)\g_i(\veq)v_{i\beta}(\veq,\vep')\dq, \nonumber
\eea
where the single thin (double) lines indicate the coupling to the open-flavour channels (pole terms) while the dashed and thick
solid lines indicate the propagation of the light $\bar{q}q$ and heavy $\bar{Q}Q$ state, respectively.

\begin{figure}[t]
\begin{center}
\epsfig{file=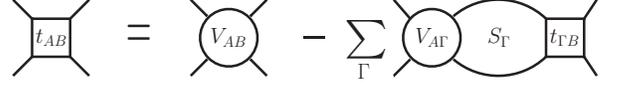,width=0.45\textwidth}
\caption{Graphical representation of the Lippmann-Schwinger equation for the elastic scattering $t$ matrix
$t_{AB}$---see Eq.~(\ref{tabnew1}).}\label{fig:lseq}
\end{center}
\end{figure}

The Lippmann-Schwinger equation for the elastic $t$ matrix $t_{AB}$ then reads (see Fig.~\ref{fig:lseq})
\bea
t_{AB}(\vep,\vep')&=&V_{AB}(\vep,\vep')\nonumber\\
&&\hspace*{-10mm}-\sum_\Gamma\int V_{A\Gamma}(\vep,\veq)\g_\Gamma(\veq)t_{\Gamma B}(\veq,\vep')\dq,
\label{tabnew1}
\eea
where $\g_\alpha(\vep)$ is the propagator of the $\alpha$-th $(\bar{q}Q)(\bar{Q}q)$ pair, and
\be
\g_{0,aa} \equiv \int \Sa(\veq)\dq=\frac{1}{M_{0,a}-M-i0}
\label{eq:s0}
\ee
denotes the nonvanishing matrix elements of the diagonal matrix of the bare pole propagators
with $M_{0,a}$ being the bare mass.
Below the results for the elastic and inelastic loop integrals will be parametrised conveniently such that the
explicit form of the propagators $\g_i(\veq)$ and $\g_\alpha(\veq)$ is of no relevance.

\begin{figure}[t]
\begin{center}
\epsfig{file=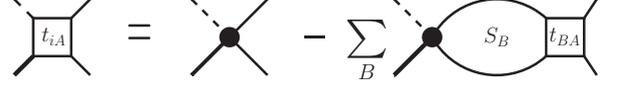,width=0.45\textwidth}
\caption{Graphical representation of the expression for the $t$ matrix component $t_{iA}$ given in
Eq.~(\ref{sysnew}). Representation for the component $t_{A i}$ takes a similar form and it is not shown.
The solid lines are for the heavy-light $(\bar{q}Q)$ and $(\bar{Q}q)$ mesons with open flavour, the fat solid line is
for
the hidden-flavour heavy meson ($\bar{Q}Q$), and the dashed line is for the light meson ($\bar{q}q$).
}\label{fig:tia}
\end{center}
\end{figure}

\begin{figure*}[t]
\begin{center}
\epsfig{file=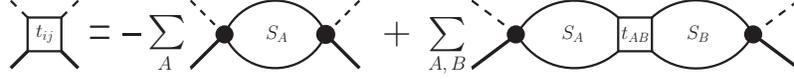,width=0.6\textwidth}
\caption{Graphical representation of the expression for the $t$ matrix component $t_{ij}$ given in
Eq.~(\ref{sysnew}). For the lines' identification see the caption of Fig.~\ref{fig:tia}.}\label{fig:tij}
\end{center}
\end{figure*}

Once a solution of the Lippmann-Schwinger equation (\ref{tabnew1}) for the elastic $t$ matrix
$t_{AB}$ is found, all other components of the $t$ matrix can be found algebraically, without
having to solve further equations (see Figs.~\ref{fig:tia} and \ref{fig:tij}),
\bea
&&\hspace*{-5mm}t_{A i}(\vep,\vek)=v_{A i}(\vep,\vek)-\sum_B\int t_{AB}(\vep,\veq)\g_B(\veq)\nonumber\\
&&\hspace*{45mm}\times v_{B i}(\veq,\vek)\dq,
\eea
\bea
&&\hspace*{-5mm}t_{iA}(\vek,\vep)=v_{iA}(\vek,\vep)-\sum_B\int v_{iB}(\vek,\veq)\g_B(\veq)\nonumber\\
&&\hspace*{45mm}t_{BA}(\veq,\vep)\dq,
\eea
\bea
&&t_{ij}(\vek,\vek')=-\sum_A\int v_{iA}(\vek,\veq)\g_A(\veq)v_{A j}(\veq,\vek')\dq\nonumber\\
&&+\sum_{A,B}\int v_{iA}(\vek,\veq)\g_A(\veq)t_{AB}(\veq,\veq')\g_B(\veq')\label{sysnew}\\
&&\hspace*{40mm}\times v_{B j}(\veq',\vek')\dq\dq'.\nonumber
\label{eq:tii}
\eea

\begin{figure}[t]
\begin{center}
\epsfig{file=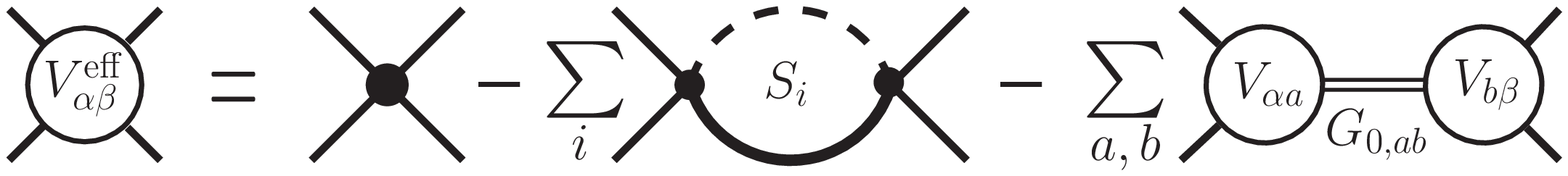,width=0.48\textwidth}
\caption{The full effective interaction potential in the elastic channels---see Eq.~(\ref{Veff}).
The double solid line is for the pole term propagator $G_0$ [see Eq.~(\ref{G0def})] while
for the other lines identification see the caption of Fig.~\ref{fig:tia}. Potentials $V_{\alpha a}$ and
$V_{b\beta}$ are defined in Eqs.~(\ref{Va0}) and (\ref{V0a}).
}\label{fig:pot}
\end{center}
\end{figure}

Equation (\ref{tabnew1}) can be written explicitly for the $t$ matrix components
$t_{\alpha\beta}$, $t_{\alpha a}$, $t_{a\alpha}$, and $t_{ab}$, and it
splits into two decoupled systems of equations. By simple algebraic transformations it
is straightforward to exclude the components $t_{ab}$ and $t_{a\alpha}$
to arrive at the following decoupled Lippmann-Schwinger equations for the
elastic $t$ matrix $t_{\alpha\beta}$,
\bea
&&\hspace*{-10mm}t_{\alpha\beta}(\vep,\vep')=V^{\rm eff}_{\alpha\beta}(\vep,\vep')\nonumber\\
&&\hspace*{10mm}-\sum_\gamma\int V^{\rm eff}_{\alpha\gamma}(\vep,\veq)\g_\gamma(\veq)t_{\gamma\beta}(\veq,\vep')\dq,
\label{tabnew}
\eea
and for the component $t_{\alpha a}(\vep)$,
\bea
\hspace*{-5mm}t_{\alpha a}(\vep)=V^{\rm eff}_{\alpha a}(\vep){-}\sum_\beta\int
V^{\rm eff}_{\alpha\beta}(\vep,\veq)\g_\beta(\veq)t_{\beta a}(\veq)\dq.
\label{ta0}
\eea
The matrix for the pole propagators dressed by the inelastic channels reads
\be
G_0=\left(\g_0^{-1}-\calg_0^\text{in}\right)^{-1},
\label{G0def}
\ee
where the inelastic loop matrix ${\cal G}_0^\text{in}$ is defined in Eq.~(\ref{V00def}).

In the single-pole case ($a=0$) $G_0$ is simply
\be
G_0=\frac{1}{M_{0}-M+V_{00}-i0}.
\label{G0def-1}
\ee
The real part of ${\cal G}_0^\text{in}=-V_{00}$
can be absorbed into the renormalisation of the bare pole position $M_0$, while its imaginary part shifts the
pole to the complex plane, away from the real axis. Note that the explicit form of  ${\cal G}_0^\text{in}$ links
the imaginary part of the pole location to the corresponding transitions to the inelastic channels as demanded
by unitarity.

For multiple bare poles the real parts of the diagonal elements
$\calg_{0,aa}^\text{in}$ can also be absorbed by the bare masses
$M_{0,a}$---see Eq.~(\ref{eq:s0})---while the off-diagonal elements
$\calg_{0,ab}^\text{in}$ ($a\neq b$) describe the transition potentials between
the bare states, and  in general their real parts,
\be
\kappa_{ab}^{\rm
in}\equiv\text{Re}(\calg_{0,ab}^\text{in}), \quad \text{for}~a\neq b,
\label{eq:kappain}
\ee
need to be retained as
additional parameters of the model. For example, in the case of two bare
poles there is one such additional parameter $\kappa_{12}^{\rm in}=\kappa_{21}^{\rm in}$.

The effective potential $V^{\rm eff}_{\alpha\beta}(\vep,\vep')$ is depicted schematically in Fig.~\ref{fig:pot}
and reads
\bea
V^{\rm eff}_{\alpha\beta}(\vep,\vep')=&&v_{\alpha\beta}(\vep,\vep')
- \sum_{a,b}V_{\alpha a}(\vep)G_{0,ab}V_{b\beta}(\vep')
\nonumber\\
&&\hspace*{-5mm}-\sum_i\int v_{\alpha
i}(\vep,\veq)\g_i(\veq)v_{i\beta}(\veq,\vep')\dq,
\label{Veff}
\eea
while the effective potential
$V^{\rm eff}_{\alpha a}$ is
\be
V_{\alpha a}^{\rm eff}(\vep)=(M_{0,a}-M)\sum_b V_{\alpha b}(\vep)G_{0,ba}.
\ee

\section{Analytic solution for separable interactions}

To proceed towards an analytic solution we assume the vertex in Eq.~(\ref{v}) to possess a separable form,
\be
v_{\alpha i}(\vep,\vek)=\chi_{i\alpha}(\vep)\varphi_{i\alpha}(\vek),
\label{vertdef}
\ee
which is necessary to express all the $t$ matrix elements in terms of those for the direct interaction, given by
Eq.~(\ref{LS0}) below.
It is obvious that the definition of Eq.~(\ref{vertdef}) is invariant under the transformation
\be
\chi_{i\alpha}(\vep)\to C\chi_{i\alpha}(\vep),\quad
\varphi_{i\alpha}(\vek)\to \varphi_{i\alpha}(\vek)/C,
\ee
with an arbitrary, real constant $C$, so that without loss of generality one can set
\be
\chi_{i\alpha}(\vep=0)=1.
\label{chinorm}
\ee

A considerable simplification of Eqs.~(\ref{tabnew}) and (\ref{ta0}) can be achieved if the form factor
$\chi_{i\alpha}(\vep)$ entering vertex function (\ref{vertdef}) is assumed independent of the
inelastic channel, that is
\be
v_{\alpha i}(\vep,\vek)=\chi_\alpha(\vep)\varphi_{i\alpha}(\vek).
\label{universalchi}
\ee
In fact, it is quite natural to assume that
$\chi_{i\alpha}$ is independent of $i$, since the transition of the open-flavour
channels to the hidden-flavour channels demands the exchange of a heavy meson
and therefore it is of a short-range nature for all inelastic channels as
long as these channels are far from the thresholds of the elastic channels (so
that the exchanged heavy meson is far off shell).
By virtue of Eq.~(\ref{universalchi}) the effective potential defined in Eq.~(\ref{Veff}) reads
\bea
V_{\alpha\beta}^{\rm eff}(\vep,\vep')=v_{\alpha\beta}(\vep,\vep')
&-&\chi_\alpha(\vep)G_{\alpha\beta}\chi_\beta(\vep')\nonumber\\
&-&\sum_{a,b}V_{\alpha a}(\vep)G_{0,ab}V_{b\beta}(\vep'),~~
\label{Valbewin}
\eea
where the inelastic bubble operator $G_{\alpha\beta}$ is
\be
G_{\alpha\beta}\equiv \sum_i G_{\alpha\beta}^i=\sum_i\int \varphi_{\alpha i}(\veq)\g_i(\veq)\varphi_{i\beta}(\veq)\dq.
\label{GG}
\ee

In order to solve Eqs.~(\ref{tabnew}) and (\ref{ta0}) we proceed stepwise.
The strategy  basically represents a successive
application of the two-potential formalism~\cite{Nakano:1982bc} (see also Ref.~\cite{Hanhart:2012wi} for an application
to a physical system more closely related to the one of relevance here):
\begin{enumerate}
\item In the first step only the direct interaction $v_{\alpha\beta}(\vep,\vep')$ [the first term in
potential (\ref{Valbewin})] is retained and a convenient parametrisation is given for the corresponding direct
interaction $t$ matrix hereinafter denoted as $\tv$;
\item then the coupling to the inelastic channels is switched on [the second term in potential (\ref{Valbewin})] and a
scattering equation for the potential
\bea
&w_{\alpha\beta}(\vep,\vep')=v_{\alpha\beta}(\vep,\vep')-\chi_\alpha(\vep) G_{\alpha\beta}\chi_\beta(\vep')&
\label{Weff}
\eea
is solved, with the solution denoted as $\tw$ (notice that here the
repeated indices do not imply a resummation which is always written
explicitly in this paper);
\item finally, the coupling to the pole terms [the last term in potential
(\ref{Valbewin})] is included, in addition.
The result provides the solution to the full problem defined in Eqs.~(\ref{tabnew}) and (\ref{ta0}).
\end{enumerate}

We therefore start assuming that a solution $\tv$ of the Lippmann-Schwinger equation
\bea
\tv_{\alpha\beta}(\vep,\vep')&=&v_{\alpha\beta}(\vep,\vep')\nonumber\\[2mm]
&-&\sum_\gamma\int v_{\alpha\gamma}(\vep,\veq)\g_\gamma(\veq)\tv_{\gamma\beta}(\veq,\vep')\dq
\label{LS0}
\eea
for the bare direct interaction $v_{\alpha\beta}(\vep,\vep')$ is
known. For instance, it can be simply parametrised in a convenient form---see Refs.~\cite{Baru:2010ww,Hanhart:2011jz}
and the discussion in Sec.~\ref{dirint} below. This finalises step 1 above.

As the coupling to the inelastic channels is switched on (step 2), the bare form
factors $\chi_\alpha(\vep)$ get dressed by the elastic interactions. It is therefore
convenient to define new incoming and outgoing form factors $\psi_{\alpha\beta}(\vep)$ and
$\bar{\psi}_{\alpha\beta}(\vep)$, respectively,
``dressed'' with the direct interaction potential $v_{\alpha\beta}(\vep,\vep')$,\footnote{Notice that once the bare form factor
$\chi_\alpha(\vep)$ does
not depend on the inelastic channel, the dressed form factors $\psi_{\alpha\beta}(\vep)$ and
$\bar{\psi}_{\alpha\beta}(\vep)$ do not depend on it either.}
\bea
\psi_{\alpha\beta}(\vep)&=&\raisebox{-2.7mm}{\epsfig{file=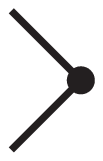,width=6mm}}-
\raisebox{-4mm}{\epsfig{file=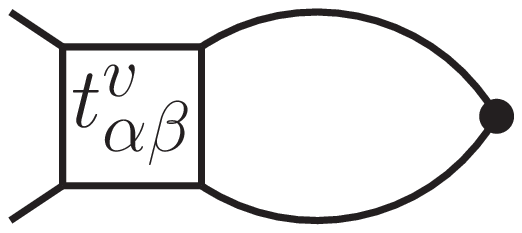,width=21mm}}\nonumber\\
&&\hspace*{-10mm}=\delta_{\alpha\beta}\chi_\alpha(\vep)-\int \tv_{\alpha\beta}(\vep,\veq)
\g_\beta(\veq)\chi_\beta(\veq)\dq,\label{psi11}\\
\bar{\psi}_{\alpha\beta}(\vep)&=&\raisebox{-2.7mm}{\epsfig{file=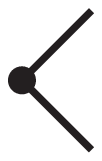,width=6mm}}-
\raisebox{-4mm}{\epsfig{file=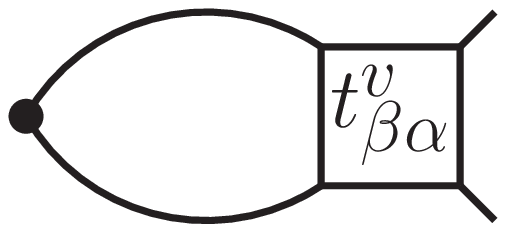,width=21mm}}\nonumber\\
&&\hspace*{-10mm}=\delta_{\alpha\beta}\chi_\alpha(\vep)-\int \chi_\alpha(\veq)
\g_\alpha(\veq)\tv_{\alpha\beta}(\veq,\vep)\dq.\label{psi12}
\eea

It is straightforward now to find the solution of the Lipp\-mann-Schwinger equation for the potential given in
Eq.~(\ref{Weff}) in the
form
\bea
\hspace*{-3mm}\tw_{\alpha\beta}(\vep,\vep')&=&\tv_{\alpha\beta}(\vep,\vep')\nonumber\\
&+&\sum_{\gamma,\delta}\psi_{\alpha\gamma}(\vep)
\left([{\cal G}-G^{-1}]^{-1}\right)_{\gamma\delta}\bar{\psi}_{\delta\beta}(\vep'),
\label{ttt}
\eea
where the matrix $G$ is given in Eq.~(\ref{GG}) while the  matrix ${\cal G}$ is defined as
\bea
{\cal G}_{\alpha\beta}&=&\raisebox{-3.5mm}{\epsfig{file=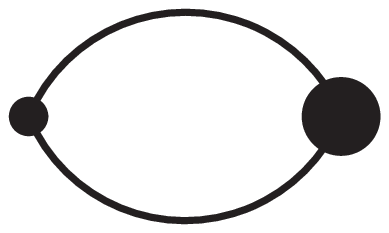,width=15mm}}
=\int\chi_{\alpha}(\veq)\g_\alpha(\veq)\psi_{\alpha\beta}(\veq)\dq\nonumber\\[-2mm]
\label{gd2}\\[-2mm]
&=&\raisebox{-3.5mm}{\epsfig{file=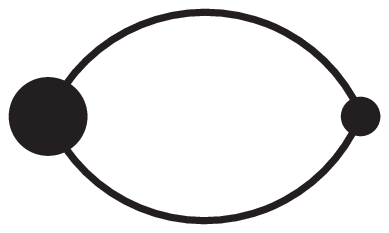,width=15mm}}
=\int\bar{\psi}_{\alpha\beta}(\veq)\g_\beta(\veq)\chi_{\beta}
(\veq)\dq.\nonumber
\eea
This finalises step 2.

To accomplish the work programme and to build solutions of Eqs.~(\ref{tabnew}) and
(\ref{ta0}) we
apply again the two-potential formalism to include the bare pole terms and to
express the full $t$ matrix elements in terms of $t^w$. This is a multi-pole
generalisation of the formulae derived in Ref.~\cite{Hanhart:2011jz}.
We find that
\bea
&&t_{\alpha\beta}(\vep,\vep')=\tw_{\alpha\beta}(\vep,\vep')\nonumber\\
&&\hspace*{20mm}+\sum_{a,b}\phi_{\alpha a}(\vep)
\left({\cal G}_0^\text{e}-G_0^{-1}\right)^{-1}_{ab}\bar{\phi}_{b\beta}(\vep'),\label{tsol}\\
&&=\tw_{\alpha\beta}(\vep,\vep')+\sum_{a,b}\phi_{\alpha a}(\vep)
\left({\cal G}_0^\text{e}+{\cal G}_0^\text{in}-S_0^{-1}\right)^{-1}_{ab}\bar{\phi}_{b\beta}(\vep'),\nonumber\\
&&t_{\alpha a}(\vep)= -\sum_b \phi_{\alpha b}(\vep)
\left[\g_0 \left({\cal G}_0^\text{e}-G_0^{-1} \right) \right]^{-1}_{ba}\nonumber\\
&&\hspace*{15mm}=\sum_b \phi_{\alpha b}(\vep)\left[1-\g_0\left({\cal G}_0^\text{e}+{\cal G}_0^\text{in}\right)\right]^{-1}_{ba},
\label{ta0sol}
\eea
where Eq.~(\ref{G0def}) was used to find $G_0^{-1}$, and
\bea
\phi_{\alpha a}(\vep) &=&
\raisebox{-2.1mm}{\epsfig{file=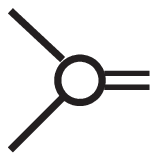,width=7mm}}-\sum_{\beta}
\raisebox{-4mm}{\epsfig{file=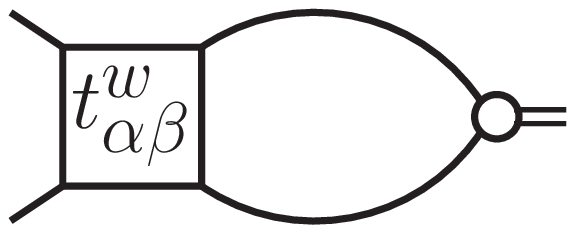,width=21mm}}\label{Phi1}\\
&=&V_{\alpha a}(\vep)-\sum_\beta\int
\tw_{\alpha\beta}(\vep,\veq)\g_\beta(\veq)V_{\beta a}(\veq)\dq,\nonumber\\
\bar{\phi}_{a\alpha}(\vep)&=&\raisebox{-2.6mm}{\epsfig{file=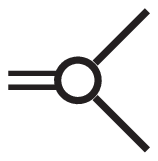,width=8mm}}-\sum_{\beta}
\raisebox{-4mm}{\epsfig{file=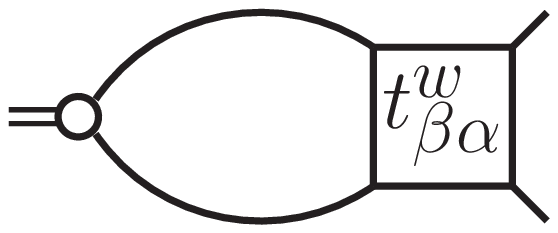,width=21mm}}\label{Phi2}\\
&=&V_{a\alpha}(\vep)-\sum_\beta\int
V_{a\beta}(\veq)\g_\beta(\veq)\tw_{\beta\alpha}(\veq,\vep)\dq,\nonumber\\
{\cal
G}_{0,ab}^\text{e}&=&\sum_\alpha\raisebox{-2.5mm}{\epsfig{file=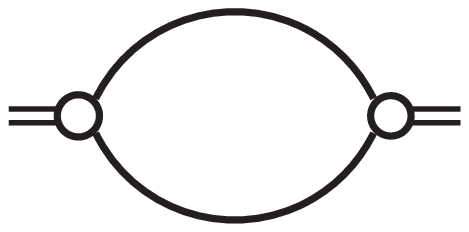,width=15mm}}-\sum_{\alpha,\beta}
\raisebox{-3.7mm}{\epsfig{file=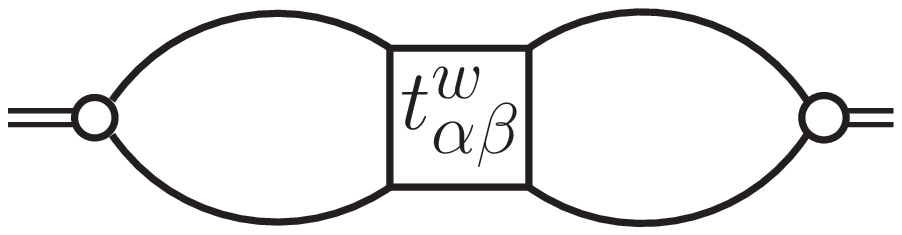,width=30mm}}\nonumber\\
&=& \sum_\alpha \int V_{a\alpha}(\veq)\g_\alpha(\veq)\phi_{\alpha b}(\veq) \dq
\nonumber\\
&=& \sum_\alpha \int \bar \phi_{a\alpha}(\veq)\g_\alpha(\veq)V_{\alpha
b}(\veq) \dq.
\label{J0}
\eea
This finalises step 3 and the entire work programme.

Finally, the $t$ matrix component $t_{\alpha i}$ can be found from the first equation in system (\ref{sysnew}),
\bea
t_{\alpha i}(\vep,\vek)&=&v_{\alpha i}(\vep,\vek)-\sum_a t_{\alpha
a}(\vep)\g_{0,aa} v_{ai}(\vek)\nonumber\\
&-&\sum_\beta\int t_{\alpha\beta}(\vep,\veq) \g_\beta(\veq)v_{\beta
i}(\veq,\vek)\dq,
\label{taisol}
\eea
so that $t_{\alpha i}$ is fully
determined through the $t$ matrices $t_{\alpha\beta}$ and $t_{\alpha a}$
explicitly found above.

The remaining components of the full $t$ matrix, namely $t_{ab}$,
$t_{a\alpha}$, $t_{i\alpha}$, and $t_{ij}$, will not be used in what follows
and therefore are not quoted here explicitly.

\section{Production amplitudes and rates}

\begin{figure}[t]
\begin{center}
\epsfig{file=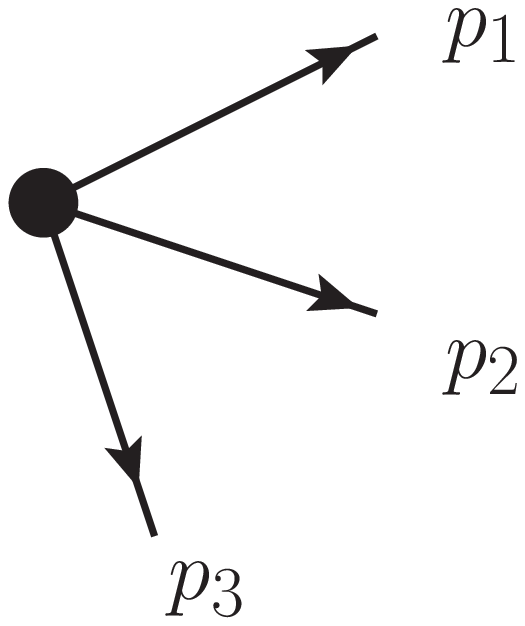,width=0.12\textwidth}\hspace*{1mm}
\raisebox{14mm}{\large $-$}
\hspace*{1mm}\raisebox{14mm}{\large $\ds\sum_\beta$}
\epsfig{file=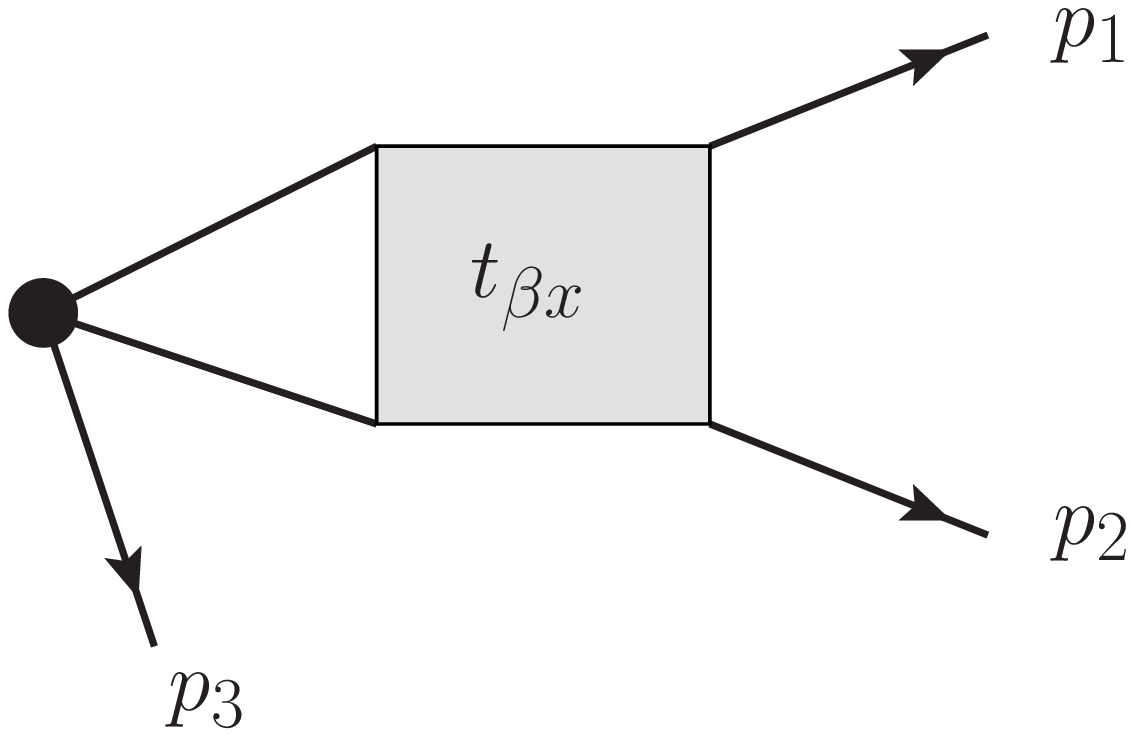,width=0.23\textwidth }
\caption{Graphical representation of the contributions to the production amplitude for the channel $x$ from a
pointlike
source: Born term (first diagram) and rescattering term (second diagram).}\label{fig:prod}
\end{center}
\end{figure}
There is no experimental possibility to study the elastic scattering of
flavoured mesons off each other, and our knowledge of the properties of
near-threshold states comes from production experiments. For the
production amplitudes one needs to add $\Ne$ sources for the elastic channels,
$\Nin$ sources for the inelastic channels and $\Np$ sources for the
pole terms. It is sufficient to treat all sources as pointlike. On the other hand,
when focusing on near-threshold phenomena it is natural to assume that the
production proceeds predominantly through the elastic channels, so that it is
sufficient to assume that only $\Ne$ elastic sources exist, with the strengths
${\cal F}_\alpha$. Therefore the production problem is set up as follows: (i) an
elastic channel is produced from a pointlike source, accompanied by a
spectator; (ii) for simplicity, the interaction in the final state between the
spectator and the other particles is neglected (this assumption allows one to
proceed with analytical calculations, but it can be relaxed in numerical
computations); (iii) due to rescatterings, any elastic or inelastic channel $x$
can be produced in the final state.

In Fig.~\ref{fig:prod} the contributions to the production amplitude in the channel $x$
(elastic or inelastic) are presented in a graphical form, and the corresponding expression reads
\be
{\cal M}^\text{e}_\alpha(\vep)={\cal F}_\alpha(\vep)-\sum_\beta\int {\cal
F}_\beta(\veq)\g_\beta(\veq)t_{\beta\alpha}(\veq,\vep)d^3q,
\label{Me}
\ee
for the elastic channel $x=\alpha$, or
\be
{\cal M}_i^\text{in}(\vek)=-\sum_\alpha \int {\cal F}_\alpha(\veq)\g_\alpha(\veq) t_{\alpha i}(\veq,\vek)d^3q,
\label{Mi}
\ee
for the inelastic channel $x=i$.

To proceed with the calculations of the differential production rates we start from the standard expression for the
differential decay width \cite{Agashe:2014kda}
\be
\frac{d\Gamma_x}{dm_{12}^2 dm_{23}^2}=\frac{1}{(2\pi)^3}\frac{1}{32M_{\rm tot}^3}\left|{\cal M}_x\right|^2,
\ee
where $M_{\rm tot}$ is the total energy of the system. Since we neglect  the final state interaction between the
spectator (particle
3) and the rest of the system (particles 1 and 2) the integration in the invariant mass $m_{23}^2$ is trivial and
yields
\bea
&&\int_{(m_{23}^2)_{\rm min}}^{(m_{23}^2)_{\rm max}}dm_{23}^2
=\frac{1}{m_{12}^2}\lambda^{1/2}(m_{12}^2,m_1^2,m_2^2)\\
&&\hspace*{15mm}\times\lambda^{1/2}(M_{\rm tot}^2,m_{12}^2,m_3^2)=\frac{4M_{\rm tot}}{m_{12}}k_{3(12)}k_{12},\nonumber
\eea
where $\lambda(x^2,y^2,z^2)$ is the standard triangle function while $k_{3(12)}\equiv p_3$
and $k_{12}\equiv k_x$ are the 3-momentum of particle 3 in the
centre-of-mass frame of particles 1 and 2, and the 3-momentum of particles 1
and 2 in the rest frame of the decaying particle, respectively. Then for the
differential rate $d\Br_x\equiv d\Gamma_x/\Gamma_{\rm tot}$ we get
\be
\frac{d\Br_x}{ds}=\frac{|{\cal M}_x|^2p_3k_x}{64\pi^3M_{\rm tot}^2\Gamma_{\rm tot}\sqrt{s}},
\label{db0}
\ee
where $s\equiv M^2=m_{12}^2$ and, consequently,
\be
\frac{d\Br_x}{dM}=\frac{|{\cal M}_x|^2p_3k_x}{32\pi^3M_{\rm tot}^2\Gamma_{\rm tot}}.
\label{db1}
\ee
Finally, the total rate comes as an integral,
\be
\Br_x=\int_{M_{\rm min}}^{M_{\rm max}}\left(\frac{d\Br_x}{dM}\right)dM,
\label{Brtot}
\ee
where $M_{\rm min}=m_1+m_2$ and $M_{\rm max}=M_{\rm tot}-m_3$.

\section{Towards a convenient parametrisation of the line shapes}
\label{sec:parametrisation-Zb}

The expressions for the $t$ matrix components and for the rates derived above can be used to build a sufficiently
general parametrisation applicable for the description of a wide class of near-threshold phenomena.

In the near-threshold region and for $S$-wave coupling of the elastic channels,
the vertex form factors $\chi_\alpha(\vep)$ can be approximated by constants
$\chi_\alpha(0)$ which, by virtue of the normalisation condition (\ref{chinorm}), are simply equal to unity. Thus, the integrals entering
Eqs.~(\ref{psi11}), (\ref{psi12}), and (\ref{gd2}) can be evaluated as
\bea
&\ds\int\chi_\alpha^2(\veq)\g_\alpha(\veq)\dq\approx\chi_\alpha^2(0)J_\alpha=J_\alpha,&\nonumber\\[-2mm]
\label{IIpr}\\[-2mm]
&\ds\int\chi_\alpha(\veq)\g_\alpha(\veq)\dq\approx\chi_\alpha(0) J_\alpha=J_\alpha,&\nonumber
\eea
where the nonrelativistic loop integral is
\be
J_\alpha=\int \g_\alpha(\vep)d^3p=(2\pi)^2\mu_\alpha(\kappa_\alpha+ik_\alpha)\equiv R_\alpha+iI_\alpha,
\label{Ja}
\ee
with $\mu_\alpha$ and $k_\alpha$ being the reduced mass and the momentum in the $\alpha$'s elastic channel,
respectively.

Then one arrives at the expressions
\bea
&\psi_{\alpha\beta}=\delta_{\alpha\beta}-\tv_{\alpha\beta}J_\beta,&\nonumber\\
&\bar{\psi}_{\alpha\beta}=\delta_{\alpha\beta}-J_\alpha\tv_{\alpha\beta},&\\
&{\cal G}_{\alpha\beta}=\delta_{\alpha\beta}J_\alpha-J_\alpha\tv_{\alpha\beta}J_\beta
\nonumber&
\eea
for the dressed form factors (\ref{psi12}) and for matrix (\ref{gd2}), respectively.

According to Eqs.~(\ref{V00def}), (\ref{eq:kappain}) and (\ref{GG}) the
contribution of the inelastic channels is given by
\bea
{\cal G}_{0,ab}^\text{in}&=& -v_{ab} + \sum_i\int
v_{ai}(\veq)\g_i(\veq)v_{ib}(\veq)\dq\label{Gcalin}\nonumber \\
&\to &\kappa_{ab}^\text{in} +\frac{i(2\pi)^2}{\sqrt{s}}\sum_i
m_{\text{th}_i^\text{in}}\mu_i^\text{in} \lambda_{ai}\lambda_{bi}
(k_i^\text{in})^{2l_i+1},~~
\label{eq:G0in}
\eea
where $\kappa_{ab}^{\rm in}$ has only off-diagonal elements (see the explanation below), and by
\bea
G_{\alpha\beta}&=&\sum_i\int \varphi_{i\alpha}(\veq)\g_i(\veq)\varphi_{i\beta}(\veq)\dq
\nonumber\\
&\to&\frac{i(2\pi)^2}{\sqrt{s}}\sum_i
m_{\text{th}_i^\text{in}}\mu_i^\text{in}
g_{i\alpha}g_{i\beta}(k_i^\text{in})^{2l_i+1},~~
\label{G0}
\eea
where the transition form factors were written in the form
\be
v_{ai}(\vek)=v_{ia}(\vek)=\lambda_{ai} |\vek|^{l_i},\quad
\varphi_{i\alpha}(\vek)=g_{i\alpha}|\vek|^{l_i},
\ee
while $l_i$, $\mu_i^\text{in}$, and $k_i^\text{in}$ are the angular
momentum, the reduced mass, and the momentum in the $i$-th inelastic channel, respectively, and
$m_{\text{th}_i^\text{in}}$ is the corresponding threshold.

In Eq.~(\ref{Gcalin}) the constant real parts
$\kappa_{ab}^{\rm in}$ include both the mixing among bare poles $v_{ab}$ and
the real parts of the inelastic loops---see Eqs.~(\ref{eq:kappain}) and
(\ref{V00def}). The diagonal elements $\kappa_{aa}^\text{in}$ should be set to zero since they
only renormalise the bare pole positions $M_{0,a}$---see the discussion above Eq.~(\ref{eq:kappain}).

Similarly, in Eq.~(\ref{G0}) the constant part of $G_{\alpha\beta}$ was omitted since it
renormalises parameters of the
direct interaction potential $v_{\alpha\beta}$---see Eq.~(\ref{Valbewin}). Equations (\ref{Gcalin}) and (\ref{G0})
provide a natural
generalisation of the K-matrix approach. Notice however that in a typical situation inelastic thresholds reside
sufficiently far below the elastic ones, so that
near the elastic thresholds, neither analyticity nor unitarity are violated by using the truncated expressions
for ${\cal G}_0^\text{in}$ and $G_{\alpha\beta}$. As was already mentioned above,
in the presented model the inelastic channels enter
additively, so that an extension of the model to include an extra inelastic channel is straightforward [see
Eqs.~(\ref{Gcalin}) and (\ref{G0})] and does not enlarge the matrices that need
to be inverted to solve the scattering problem. In the case of only remote
inelastic channels the dependence of the momenta $k_i^\text{in}$ on the energy can be neglected.
Therefore, if the open inelastic thresholds reside far away from the energy region of interest (in particular, far
from the elastic
thresholds), their contribution can be mimicked by simply giving the bare pole positions $M_{0,a}$ as well as the
direct interaction potentials $v_{ab}$ and $v_{\alpha\beta}$ a constant imaginary part.

It is straightforward now to build the $t$ matrix $\tw$ given by Eq.~(\ref{ttt}) as
\be
\tw(M)=\tv(M)+\psi[{\cal G}-G^{-1}]^{-1}\bar{\psi}.
\label{ttt2}
\ee

If the bare pole terms are present in the system then, similarly to
Eq.~(\ref{IIpr}), one can write
\bea
&\ds\int
v_{a\alpha}(\veq)\g_\alpha(\veq)
v_{\alpha b}(\veq)\dq\approx v_{a\alpha} v_{b\alpha}
J_\alpha,&\nonumber\\[-2mm] \label{ggpr}\\[-2mm]
&\ds\int v_{a\alpha}(\veq)\g_\alpha(\veq)\dq\approx v_{a\alpha}
J_\alpha,&\nonumber
\eea
where $v_{a\alpha}=v_{a\alpha}(0)=v_{\alpha a}(0)$, and
\be
V_{\alpha a}=V_{a\alpha}=
v_{a\alpha} -\frac{i(2\pi)^2}{\sqrt{s}}\sum_i
m_{\text{th}_i^\text{in}}\mu_i^\text{in}
g_{i\alpha}\lambda_{ai}(k_i^\text{in})^{2l_i+1},
\label{eq:Valphaa}
\ee
where, as before, the energy-independent parts of the sums were absorbed into the renormalisation of the constants
$v_{a\alpha}$.

Then quantities $\phi_{\alpha a}$, $\bar{\phi}_{a\alpha}$ and ${\cal G}^\text{e}_{0,ab}$ defined in Eqs.~(\ref{Phi1}), (\ref{Phi2}), and (\ref{J0})
can be built as
\bea
\phi_{\alpha a}&=&V_{\alpha a}-\sum_\beta\tw_{\alpha\beta}J_\beta V_{\beta
a},\nonumber\\
\bar{\phi}_{a\alpha}&=&V_{a\alpha}-\sum_\beta V_{a\beta}J_\beta
\tw_{\beta\alpha},\label{phis}\\
{\cal G}_{0,ab}^\text{e}&=&\sum_\alpha V_{a\alpha}J_\alpha V_{\alpha
b}-\sum_{\alpha,\beta}V_{a\alpha}J_\alpha \tw_{\alpha\beta} J_\beta V_{\beta
b},\nonumber
\eea
respectively, which when substituted into Eqs.~(\ref{tsol}), (\ref{ta0sol}) and
(\ref{taisol}) allow one to find the expressions for $t_{\alpha\beta}$,
$t_{\alpha a}$ and $t_{\alpha i}$ in their ultimate form.

Then for the $\alpha$th elastic channel in the final state and for constant sources ${\cal F}_\alpha$ production
amplitude (\ref{Me}) is
\be
{\cal M}^\text{e}_\alpha={\cal F}_\alpha-\sum_\beta\int {\cal F}_\beta\g_\beta(\veq)t_{\beta\alpha}\dq
={\cal F}_\alpha-\sum_{\beta}{\cal F}_\beta J_\beta t_{\beta\alpha}.
\label{eq-am}
\ee
If the $t$ matrix has near-threshold poles, then the Born term can be neglected,
provided that we focus on the near-threshold region (a detailed
discussion can be found in Ref.~\cite{Baru:2010ww}).
Strictly speaking, neglecting the Born term violates unitarity; however this violation is negligibly small and it is
controlled by the proximity of the $t$ matrix poles to the threshold(s).\footnote{In certain cases,
however, the Born term can play a crucial role as discussed, e.g., in Ref.~\cite{Guo:2014iya}.}

Similarly, for the $i$th inelastic channel in the final state we have [see Eq.~(\ref{Mi})]
\be
{\cal M}_i^\text{in}=-\sum_\alpha {\cal F}_\alpha J_\alpha t_{\alpha i}.
\label{eq-bm}
\ee

Accordingly the expressions for the differential production rates are
\be
\frac{d\Br_\alpha}{dM}=\Bigl|\sum_\beta {\cal F}_\beta t_{\beta\alpha}\Bigr|^2p_3 k_\alpha
\label{Bre}
\ee
and
\be
\frac{d\Br_i}{dM}=\Bigl|\sum_\alpha{\cal F}_\alpha t_{\alpha i}\Bigr|^2p_3k_i^\text{in},
\label{Bri}
\ee
where the source terms ${\cal F}_\alpha$ were redefined to
absorb the slowly varying function of energy $J_\alpha=R_\alpha+iI_\alpha\approx R_\alpha$ as well
as all constant factors from Eq.~(\ref{db1}).

To simplify notations and to make the physical meaning of the parameters more transparent we define
\be
{\cal N}={\cal F}_1^2,\quad \xi_{\alpha}={\cal F}_{\alpha}/{\cal F}_1.
\label{eq:num}
\ee
In addition, since for all elastic channels the range of forces is described by the same physics, it is natural to
set
$\kappa_\alpha=\kappa$ in all $R_\alpha$'s [see Eq.~(\ref{Ja})].

Therefore, the line shapes for the production in $\Ne$ elastic and $\Nin$ inelastic channels are described by the
following set of parameters:
\be
{\cal N},~\xi_\alpha,~v_{a\alpha},~\lambda_{ai},~g_{i\alpha},~M_{0,a},~\kappa,~\kappa_{ab}^{\rm
in} (a\neq
b),~\tv,
\label{set1}
\ee
that is by $\Nv+\Nin(\Ne+\Np)+(\Np+1)(\Ne+1)+\Np(\Np-1)/2+1$ real parameters
($\Nv$ is the number of parameters for the direct interaction $t$ matrix $\tv$). Notice that the constants $v_{ab}$ are not independent parameters
since they were included into the definition of $\calg_{0,ab}^\text{in}$ and thus they are absorbed by $\kappa_{ab}^\text{in}$---see
Eqs.~(\ref{V00def}) and (\ref{Gcalin}). The number of parameters can be reduced if the analysed system possesses a symmetry which constrains some of
the parameters from Eq.~(\ref{set1}).

Then for the elastic and inelastic differential rates one finally finds:
\bea
&\ds\frac{d\Br_\alpha^\text{e}}{dM}={\cal N} \Bigl|\sum_\beta \xi_\beta t_{\beta\alpha}\Bigr|^2p_3k_\alpha,&\label{BreBB0}\\
&{\ds\frac{d\Br_i^\text{in}}{dM}={\cal N} \Bigl|\sum_\alpha \xi_\alpha
t_{\alpha i}\Bigr|^2p_3k^\text{in}_i}.&\label{Brin0}
\eea
In order to arrive at the final expressions various momentum dependencies that are suppressed
kinematically in the near-threshold regime were dropped. We confirmed the applicability of those approximations by comparing
the analytic solution presented above with a solution of the full equations found numerically.

\section{Direct interaction in the $(\bar{q}Q)(\bar{Q}q)$ system}
\label{sec:direct}

A paradigmatic example of a near-threshold state des\-crib\-ed by the general formulae derived in the previous section
(in fact by their simplified version given by the two-channel Flatt{\'e} distribution) is the glorious $X(3872)$
charmonium((like)) state discovered by the Belle Collaboration in 2003 \cite{Choi:2003ue} which resides within less than
1~MeV from
the neutral $D\bar{D}^*$ thre\-shold \cite{Agashe:2014kda}. There exists a vast literature on the description of the $X$
line
shapes in its open-charm and hidden-charm decay channels---see, for example,
Refs.~\cite{Hanhart:2007yq,Kalashnikova:2009gt,Kalashnikova:2010zz,Braaten:2007dw,Stapleton:2009ey,Artoisenet:2010va,Hanhart:2007yq,Danilkin:2010cc}
to mention just a few. We
therefore do not dwell on the $X(3872)$ any more and consider another intriguing example of near-threshold phenomena
provided by the $Z_b^{(\prime)}$ resonances discovered by the Belle Collaboration in 2011 in the spectrum of bottomoniumlike states
\cite{Belle:2011aa} and which appear as intermediate states in the $\Upsilon(5S)$ decays
\cite{Bondar:2011ev,Voloshin:2011qa}. Proximity of the observed $Z_b(10610)$ and $Z_b(10650)$ to the
$B\bar{B}^*$ and $B^*\bar{B}^*$ thresholds, respectively hints towards a prominent molecular component
of both states \cite{Bondar:2011ev} and
calls for a simultaneous description of the available experimental data for their open- and hidden-bottom decay
channels.\footnote{It was shown recently that the $Z_b$ states even play a
crucial role in understanding the transitions
$\Upsilon(3S)\to\Upsilon(1S)\pi\pi$~\cite{Chen:2015jgl}.}

\subsection{Contact elastic interaction potential}

The four negative-parity heavy-light $B$ mesons have the wave functions
\be
B=0^-_{\bar{b}q},\quad\bar{B}=0^-_{\bar{q}b},\quad B^*=1^-_{\bar{b}q},\quad\bar{B}^*=-1^-_{\bar{q}b},
\label{CB}
\ee
where, for example, the symbol $0^-_{\bar{b}q}$ denotes the quantum numbers $J^P=0^-$ in the system of antiquark $\bar{b}$ and the light quark $q$.
The charge conjugation operation for a meson ${\cal M}$ is defined as
\be
{\hat C}{\cal M}=\bar{\cal M}.
\label{Cpardef}
\ee

The direct interaction potential in the elastic channels can be extracted from the effective Lagrangian
which describes the $B^{(*)}\bar{B}^{(*)}$ interactions consistent with the
heavy-quark spin symmetry (HQSS)~\cite{AlFiky:2005jd,Nieves:2012tt}.
Alternatively, if the source of the interaction in the $B^{(*)}\bar{B}^{(*)}$ channels is identified with $u$-channel
quark exchanges then the problem reduces to
performing a Fierz transformation from the open-bot\-tom states $(J_{\bar{q}b}^-\otimes
J_{\bar{b}q}^-)_S$ to the hidden-bottom states $(J_{\bar{b}b}^-\otimes J_{\bar{q}q}^-)_S$~\cite{Bondar:2011ev,Voloshin:2011qa},
\bea
&&(0^-_{\bar{q}b}\otimes 0^-_{\bar{b}q})_{S=0}=\frac{1}{2}(0^-_{\bar{b}b}\otimes 0^-_{q \bar
q})_{S=0}\nonumber\\
&&\hspace*{40mm}-\frac{\sqrt{3}}{2}(1^-_{\bar{b}b}\otimes 1^-_{\bar{q}q})_{S=0},
\label{scalar1}
\eea
\bea
&&(1^-_{\bar{q}b}\otimes 1^-_{\bar{b}q})_{S=0}=-\frac{\sqrt{3}}{2}(0^-_{\bar{b}b}\otimes 0^-_{\bar{q}q})_{S=0}\nonumber\\
&&\hspace*{40mm}-\frac{1}{2}(1^-_{\bar{b}b}\otimes 1^-_{\bar{q}q})_{S=0},
\label{scalar2}
\eea
\bea
&&(1^-_{\bar{q}b}\otimes 0^-_{\bar{b}q})_{S=1}=\frac{1}{2}(1^-_{\bar{b}b}\otimes 0^-_{\bar{q}q})_{S=1}+\frac{1}{2}
(0^-_{\bar{b}b}\otimes 1^-_{\bar{q}q})_{S=1}\nonumber\\
&&\hspace*{40mm}-\frac{1}{\sqrt{2}}(1^-_{\bar{b}b}\otimes 1^-_{\bar{q}q})_{S=1},
\label{axial1}
\eea
\bea
&&(0^-_{\bar{q}b}\otimes 1^-_{\bar{b}q})_{S=1}=\frac{1}{2}(1^-_{\bar{b}b}\otimes 0^-_{\bar{q}q})_{S=1}
+\frac{1}{2}(0^-_{\bar{b}b}\otimes 1^-_{\bar{q}q})_{S=1}\nonumber\\
&&\hspace*{40mm}+\frac{1}{\sqrt{2}}(1^-_{\bar{b}b}\otimes 1^-_{\bar{q}q})_{S=1},
\label{axial2}
\eea
\bea
&&(1^-_{\bar{q}b}\otimes 1^-_{\bar{b}q})_{S=1}=-\frac{1}{\sqrt{2}}(1^-_{\bar{b}b}\otimes 0^-_{\bar{q}q})_{S=1}\nonumber\\
&&\hspace*{40mm}+\frac{1}{\sqrt{2}}(0^-_{\bar{b}b}\otimes 1^-_{\bar{q}q})_{S=1},
\label{axial3}
\eea
\bea
&&(1^-_{\bar{q}b}\otimes 1^-_{\bar{b}q})_{S=2}=(1^-_{\bar{b}b}\otimes 1^-_{\bar{q}q})_{S=2}.
\label{tensor}
\eea

Since, in the heavy-quark limit, the transition potential in the elastic channels depends only on the light degrees of
freedom, then only two parameters (potentials) are needed:
\be
V[0^-_{\bar{q}q}]\equiv \VI,\quad  V[1^-_{\bar{q}q}]\equiv \VII.
\label{VVV}
\ee
With the help of Eqs.~(\ref{scalar1})--(\ref{tensor}) it is straightforward to find for
the transition potentials in various channels:
\be
v(0^{++})=\frac14
\left(
\begin{array}{cc}
\VI+3\VII&\sqrt{3}(\VI-\VII)\\[3mm]
\sqrt{3}(\VI-\VII)&3\VI+\VII
\end{array}
\right),
\label{0pp}
\ee
\be
v(1^{+-})=\frac12
\left(
\begin{array}{cc}
\VI+\VII&\VII-\VI\\[3mm]
\VII-\VI&\VI+\VII
\end{array}
\right),
\label{1pm}
\ee
\be
v(1^{++})={}_{1^{++}}\langle B\bar{B}^*|\hat{V}(1^{++})|B\bar{B}^*\rangle_{1^{++}}=\VII,
\label{1pp}
\ee
\be
v(2^{++})={}_{2^{++}}\langle B^*\bar{B}^*|\hat{V}(2^{++})| B^*\bar{B}^*\rangle_{2^{++}}=\VII,
\label{2pp}
\ee
where in Eq.~(\ref{1pm}) it was used that, according to Eq.~(\ref{Cpardef}), the $C$-odd combinations of the
$B^{(*)}$ and $B^*$ mesons are \cite{Bondar:2011ev,Voloshin:2011qa}
\bea
&&\hspace*{-2mm}|B\bar{B}^*\rangle_{1^{+-}}=\frac{1}{\sqrt{2}}(|B\bar{B}^*\rangle{-}|\bar{B}B^*\rangle)\label{eq:Zb}\\
&&\hspace*{15mm}=-\frac{1}{\sqrt 2}\Bigl[(1^-_{\bar{b}b}\otimes 0^-_{\bar{q}q})_{S=1}+
(0^-_{\bar{b}b}\otimes 1^-_{\bar{q}q})_{S=1}\Bigr]\nonumber,\\
&&\hspace*{-2mm}|B^*\bar{B}^*\rangle_{1^{+-}}=\frac{1}{\sqrt{2}}\Bigl[(1^-_{\bar{b}b}\otimes 0^-_{\bar{q}q})_{S=1}-(0^-_{\bar{b}b}\otimes
1^-_{\bar{q}q})_{S=1}\Bigr].\label{eq:Zbp}
\eea

The transition potentials of Eqs.~(\ref{0pp})--(\ref{2pp}) are equivalent to those obtained in
Ref.~\cite{Nieves:2012tt} [Eqs.~(18)--(21)]. To recover the latter one is to redefine
the contact potentials
\be
C_{0a}=\frac{1}{4}\VI+\frac{3}{4}\VII,\quad C_{0b}=-\frac{1}{4}\VI+\frac{1}{4}\VII
\label{Cs}
\ee
and to stick to a different definition of the $C$-parity used in Ref.~\cite{Nieves:2012tt} that eventually only entails
a change of the signs of the off-diagonal terms in the potential $v(0^{++})$.

\subsection{Direct interaction $t$ matrix}\label{dirint}

For a given momentum-independent direct interaction potential $v_{\alpha\beta}$ the $t$ matrix $\tv$ can be found from
Eq.~(\ref{LS0}),
\be
\tv_{\alpha\beta}=v_{\alpha\beta}-\sum_\gamma v_{\alpha\gamma} J_\gamma \tv_{\gamma\beta},
\label{LSE}
\ee
where the loop integrals
$J_\alpha$ are defined in Eq.~(\ref{Ja}) above.
The solution of Eq.~(\ref{LSE}) then reads
\be
(\tv)^{-1}=v^{-1}+(R+iI)=v_{\rm ren}^{-1}+iI,
\label{ren}
\ee
where the real part of the loop operator $R$ is absorbed into the renormalisation of the contact potential $v$  as
\be
v_{\rm ren}=Z^{-1}v,\quad Z=1+v R.
\ee
Since the direct interaction potential is an input for the model, it is sufficient to stick to its renormalised value
from the beginning and therefore the subscript ``ren'' can be dropped. In addition, this justifies omitting in
Eq.~(\ref{LSE}) all real parts of the loops defined in Eq.~(\ref{Ja}).

For the channels $1^{++}$ and $2^{++}$ Eq.~(\ref{LSE}) reduces
to a single equation $\tv=v-vI\tv$ with the solution
\be
\tv=\frac{1}{(2\pi)^2\mu}(\gamma_V+i k)^{-1},\quad \gamma_V^{-1}=(2\pi)^2\mu v,
\label{tvv}
\ee
where, as was explained above, the real part of the loop integral $J=\int \g(\veq)\dq=R+iI$ is absorbed into the
potential $v$ while its
imaginary part $(2\pi)^2\mu k$ is retained explicitly. Here $\mu$ and $k$ are the reduced mass and the momentum
in the corresponding $B^*\bar{B}^{(*)}$ system, respectively.

For the channels $0^{++}$ and $1^{+-}$ Eq.~(\ref{LSE}) turns into a system of two coupled
equations with the solution
\be
\tv=\frac{1}{\Delta}
\left(
\begin{array}{cc}
v_{11}+\Delta_v J_2&v_{12}\\
v_{21}&v_{22}+\Delta_v J_1
\end{array}
\right),
\label{tvVV}
\ee
where
\bea
&\Delta_v=v_{11}v_{22}-v_{12}v_{21},&\\[2mm]
&\Delta=1+v_{11}J_1+v_{22}J_2+\Delta_v J_1J_2.& \label{eq:det}
\eea
As before, the real parts of the loop integrals $J_\alpha$ can be absorbed into a redefinition of the potential
$v_{\alpha\beta}$. The quantities $\mu_\alpha$ and $k_\alpha$ are the reduced mass and the momentum in the
$B^{(*)}\bar{B}^{(*)}$ channel $\alpha$, respectively. In the nonrelativistic limit
\bea
k_\alpha(E)&=&\sqrt{2\mu_\alpha(E-\Delta_\alpha)}\Theta(E-\Delta_\alpha)\nonumber\\
&+&i\sqrt{2\mu_\alpha(\Delta_\alpha-E)}
\Theta(\Delta_\alpha-E),
\label{kE}
\eea
where $\Delta_\alpha$ is the position of the corresponding elastic thre\-shold and the energy is conveniently counted
from the lowest of them, $E=M-m_{th}$.

For the quantum numbers $1^{+-}$, relevant for the $Z_b^{(\prime)}$'s case [see Eq.~(\ref{1pm})],
\bea
&&v_{11}=v_{22}=\frac12(\VI+\VII),\nonumber\\[-2mm]
\label{vs}\\[-3mm]
&&v_{12}=v_{21}=\frac12(\VII-\VI). \nonumber
\eea
It is convenient then to introduce parameters $\gamma_s$ and $\gamma_t$ such that
\bea
\gamma_s^{-1}&=&(2\pi)^2\mu(v_{11}+v_{12})=(2\pi)^2\mu \VII,\nonumber\\[-2mm]
\label{gammast}\\[-2mm]
\gamma_t^{-1}&=&(2\pi)^2\mu(v_{11}-v_{12})=(2\pi)^2\mu \VI,\nonumber
\eea
where, for simplicity, the difference between the reduced masses in the channels $B\bar{B}^*$ and $B^*\bar{B}^*$ is
neglected, so that $\mu_1=\mu_2=\mu$.

When expressed in terms of the new parameters $\gamma_s$ and $\gamma_t$, the direct interaction $t$ matrix given by
Eq.~(\ref{tvVV}) takes the form:
\be
t^v=\frac{1}{(2\pi)^2\mu}\frac{1}{\mbox{Det}}
\left(
\begin{array}{cc}
\frac12(\gamma_s+\gamma_t)+ik_2&\frac12(\gamma_t-\gamma_s)\\[1mm]
\frac12(\gamma_t-\gamma_s)&\frac12(\gamma_s+\gamma_t)+ik_1
\end{array}
\right),
\label{tv}
\ee
with
\be
\mbox{Det}=\gamma_s\gamma_t-k_1k_2+\frac{i}{2}(\gamma_s+\gamma_t)(k_1+k_2).
\label{det}
\ee

\section{Line shapes of the $Z_b$ and $Z_b'$}

\begin{table*}
\begin{ruledtabular}
\begin{tabular}{cccccccc}
Fit& Data& $\gamma_s$, MeV & $\gamma_t$, MeV &  $\xi$
& $g_{[\pi h_b(1P)][B^*\bar{B}^*]}/g_{[\pi h_b(1P)][B\bar{B}^*]}$ &
$g_{[\pi h_b(2P)][B^*\bar{B}^*]}/g_{[\pi h_b(2P)][B\bar{B}^*]}$&C.L.\\
\hline
A& Old& $-39\pm 11$&$-137\pm 29$& $-1$&1&1& 32\%\\
\hline
B& New& $-70^{+32}_{-36}$&$-83^{+35}_{-38}$& $-1$&1&1& 48\%\\
\hline
C& New& $43^{+37}_{-58}$&$-211^{+68}_{-58}$& $-0.80\pm0.10$&$1.8^{+0.9}_{-0.5}$&$1.8^{+0.9}_{-0.5}$& 53\%
\end{tabular}
\end{ruledtabular}
\caption{Parameters of the model determined from the combined fit to the data for the
$\pi h_b(mP)$ final state contained in Ref.~\cite{Belle:2011aa} and for the $B^{(*)}\bar{B}^*$ final state contained in
Ref.~\cite{Adachi:2012cx} (denoted as old data) and in Ref.~\cite{Garmash:2015rfd} (denoted as new
data).}\label{tab:params}
\end{table*}

To exemplify the potential of the parametrisation derived in this paper we use the
latter to describe the line shapes of the $Z_b(10610)$ and $Z_b(10650)$
bottomoniumlike states. For other discussions on the line shapes of the $Z_b$'s, we
refer to Refs.~\cite{Cleven:2011gp,Mehen:2013mva,Huo:2015uka}. We consider the
simplest possible version of the formulae thus refraining from inclusion of the
bare poles that corresponds to setting $v_{a\alpha}(\vep)=v_{ai}(\vek)=0$ and
$M_{0,a}\to\infty$ in all formulae above. It should be noticed that
inclusion of one or two explicit poles would result in a fit of comparable quality.
However, since the data can already be very well described without bare poles, such a
fit would not be better and the couplings for the bare states would get little
constrained. Thus, at the present stage and given the quality of the data currently
available, the bare pole terms are not needed.

The existing experimental data for the $Z_b$'s are exhausted by 7 decay chains:
\bea
\Upsilon(5S)&\to&\pi Z_b^{(\prime)}\to \pi B^{(*)}\bar{B}^*,\nonumber\\
\Upsilon(5S)&\to&\pi Z_b^{(\prime)}\to\pi\pi\Upsilon(nS),\quad n=1,2,3,\label{chains}\\
\Upsilon(5S)&\to&\pi Z_b^{(\prime)}\to\pi\pi h_b(mP),\quad m=1,2.\nonumber
\eea

Therefore, in the formulae derived above the spectator particle is the pion (particle 3 in Fig.~\ref{fig:prod}) and,
with the help of Eqs.~(\ref{BreBB0}) and (\ref{Brin0}), we find for
the production rates in two elastic [$B\bar{B}^*$ and $B^*\bar{B}^*$] and five inelastic [$\pi\Upsilon(nS)$ and
$\pi h_b(mP)$] channels
\bea
&\ds\frac{d\Br_1^\text{e}}{dM}={\cal N}\Bigl|t_{11}+\xi t_{21}\Bigr|^2p_\pi k_1,&\nonumber\\
&\ds\frac{d\Br_2^\text{e}}{dM}={\cal N}\Bigl|t_{12}+\xi t_{22}\Bigr|^2p_\pi k_2,&\label{Brei}\\
&\ds\frac{d\Br_i^\text{in}}{dM}={\cal N} R^2\Bigl|g_{i1}(t_{11}+\xi t_{21})+g_{i2}(t_{12}+\xi
t_{22})\Bigr|^2&\nonumber\\
&\hspace*{60mm}\times p_\pi(k^\text{in}_i)^{2l_i+1},&\nonumber
\eea
respectively, where $l_i$ is the angular momentum in the final state.
Analysis of the angular distributions favours the $J^P=1^+$
assignment for both $Z_b$ states \cite{Collaboration:2011gja}. Since
the structures of interest are very close to the $B\bar{B}^*$ and $B^*\bar{B}^*$ thresholds, in the analysis we only take into account the lowest
possible orbital angular momenta for the coupled channels, which are the $S$ wave for the $B\bar{B}^*$, $B^*\bar{B}^*$, and $\pi\Upsilon(nS)$ channels
and the $P$ wave for the $\pi h_b(mP)$ channels. Therefore, in Eq.~(\ref{Brei}) above, $l_i=0$ for the $\pi\Upsilon(nS)$ channels and $l_i=1$ for the 
$\pi
h_b(mP)$ ones while $t_{11}$, $t_{12}$, $t_{21}$, $t_{22}$ are the components of the $2\times 2$ elastic $t$ matrix 
$t_{\alpha\beta}$.
As was explained above [see Eq.~(\ref{eq:num})], instead of the original quantities ${\cal F}_1$ and ${\cal F}_2$ we
introduced the overall normalisation parameter ${\cal N}$ and the ratio $\xi$ and, for simplicity, set
$\mu_1=\mu_2\equiv\mu$ so that the quantity $\kappa$ is defined as $R_1=R_2\equiv R=(2\pi)^2\mu\kappa$.

Two comments are in order here. First, as was explained before, we
neglect the $\pi\pi$ interaction in the final state although it would be needed to
ensure exact three-body unitarity. However, since the aim of the suggested approach is
to fit the structures in the $\pi \Upsilon(nS)$ and $\pi h_b(mP)$ invariant mass
distributions, the cross-channel $\pi\pi$ interaction can only provide a smooth
background. In particular, we do not expect the $\pi\pi$ interaction to produce narrow
structures in the studied channels. Therefore, while being important when it comes to
fitting the two-pion invariant mass distributions in the $\pi\pi \Upsilon(nS)$
channels, the $\pi\pi$ final state interaction is not expected to have any significant
impact on the observables discussed in this paper.

The other comment is that, in addition to the three-body pointlike source
terms $\Upsilon(5S)\to B^{(*)}\bar{B}^*\pi$ which correspond to the black dot in
Fig.~\ref{fig:prod}, the pion emission may proceed from the $B$-meson lines. Such processes were studied in detail
in Ref.~\cite{Mehen:2013mva} and it can be concluded from the results reported there that, at the tree level, the
amplitude with such a sequential pion emission is strongly suppressed compared to the three-body pointlike source
term. We therefore disregard them here and treat the production mechanism depicted in Fig.~\ref{fig:prod} as the
dominating mechanism.

According to Eq.~(\ref{tv}) the direct interaction elastic $t$ matrix $\tv$ is parametrised with 2 parameters
$\gamma_s$ and $\gamma_t$ and therefore we arrive at the following set of 15 parameters describing the line shapes in 7
elastic and inelastic channels for the $Z_b$'s [see Eq.~(\ref{chains}]
\be
\gamma_s,~\gamma_t,~\kappa,~\xi,~{\cal N},~g_{i\alpha},
\ee
where $i=\pi\Upsilon(nS)$, $\pi h_b(mP)$ with $n=1,2,3$, $m=1,2$ and $\alpha=B\bar{B}^*$, $B^*\bar{B}^*$.

We perform a simultaneous fit for the background-subtracted and efficiency-corrected distributions in $M$ for the $B^{(*)}\bar{B}^*$
\cite{Adachi:2012cx,Garmash:2015rfd} and $\pi h_b(mP)$ channels~\cite{Belle:2011aa}. We cannot fit line shapes in the $\pi\Upsilon(nS)$ channels
since they have a significant nonresonant contribution that depends on $M(\pi\pi)$; thus the amplitude analysis has to be multidimensional.
Instead, we can predict the $Z_b^{(\prime)}$ line shapes in these channels, as discussed below. Normalisations in different channels are floated
independently and we use the measured production cross sections of all seven
channels~\cite{Adachi:2012cx,Garmash:2015rfd,Belle:2011aa,Adachi:2011ji,Garmash:2014dhx,Abdesselam:2015zza} as additional constraints to ensure the 
correct
relative probabilities for the analysed distributions. The finite experimental
resolution is accounted for via a convolution of the resulting distributions with a Gaussian with $\sigma=6$~MeV. Since $\kappa$ is practically
unconstrained by the fit we fix it to 1~GeV.

As was explained above, the number of parameters can be reduced if some symmetry constraints are applied. In
particular, for the system at hand HQSS constraints following from Eqs.~(\ref{eq:Zb}) and (\ref{eq:Zbp}) read
\be
\frac{g_{[\pi\Upsilon(nS)][B^*\bar{B}^*]}}{g_{[\pi\Upsilon(nS)][B\bar{B}^*]}}=-1,\quad
\frac{g_{[\pi h_b(mP)][B^*\bar{B}^*]}}{g_{[\pi h_b(mP)][B\bar{B}^*]}}=1,
\label{HSconstr}
\ee
where $n=1,2,3$ and $m=1,2$. In addition, as the elastic channels $B\bar{B}^*$ and $B^*\bar{B}^*$ are
produced in the decays
of the $\Upsilon(5S)$ bottomonium [see Eq.~(\ref{chains})], then the ratio of the sources $\xi$ is subject to
the same heavy-quark constraint, that is
\be
\xi=\frac{g_{[\pi\Upsilon(5S)][B^*\bar{B}^*]}}{g_{[\pi\Upsilon(5S)][B\bar{B}^*]}}=-1.
\label{HSconstrxi}
\ee

\begin{figure*}[t]
\begin{center}
\epsfig{file=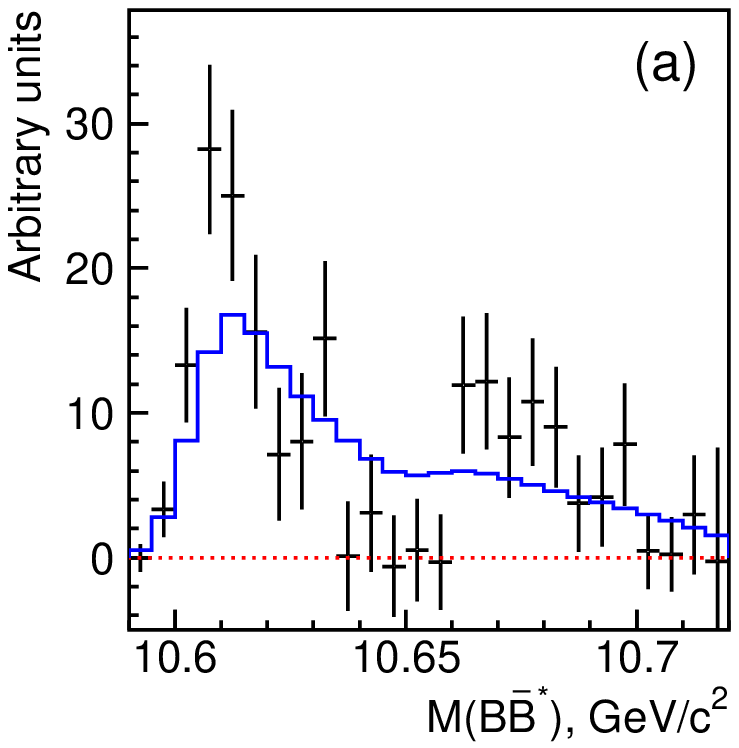,width=0.24\textwidth}
\epsfig{file=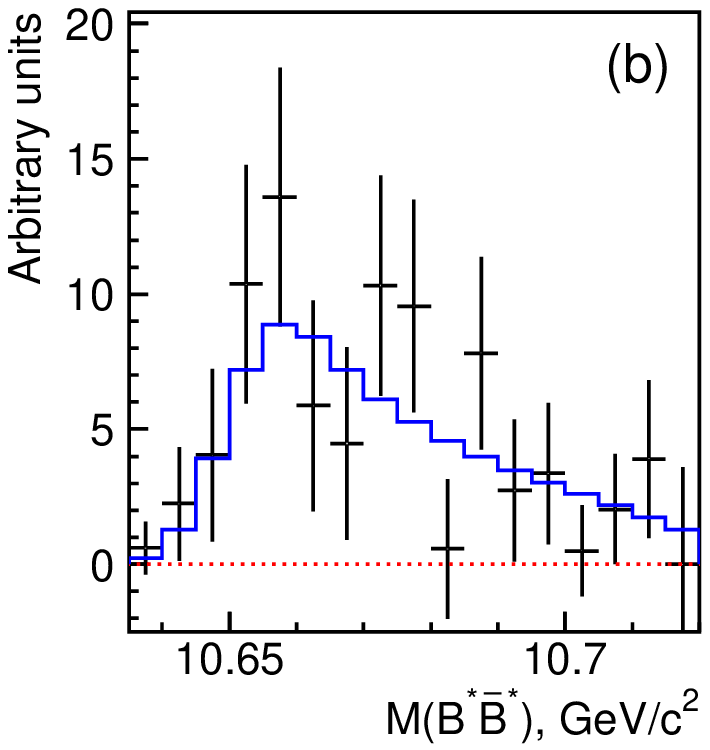,width=0.24\textwidth}
\epsfig{file=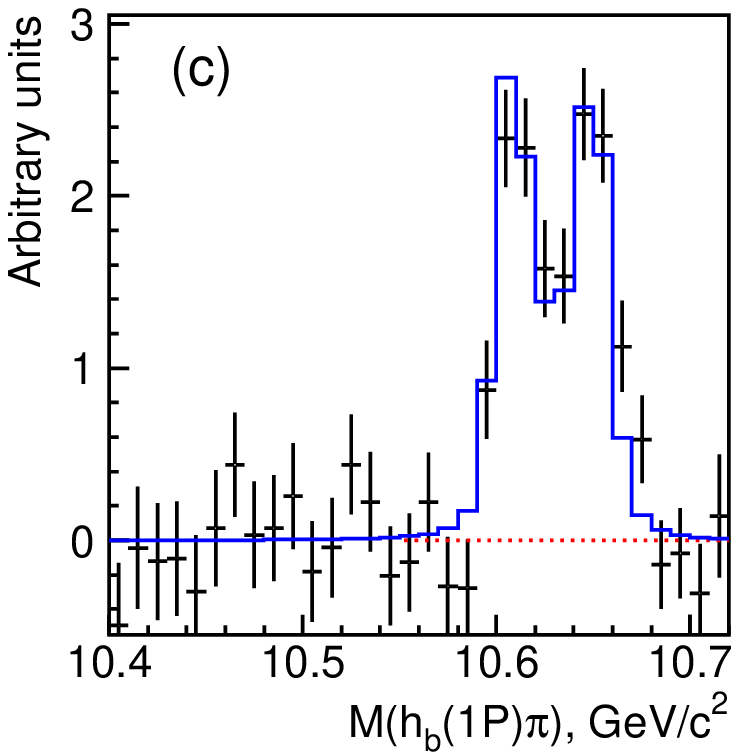,width=0.24\textwidth}
\epsfig{file=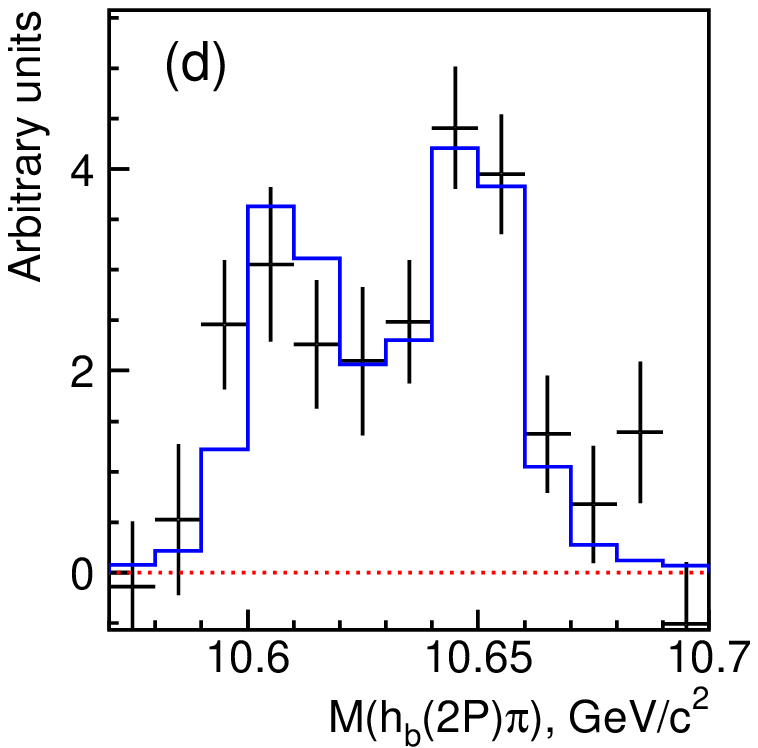,width=0.24\textwidth}\\
\epsfig{file=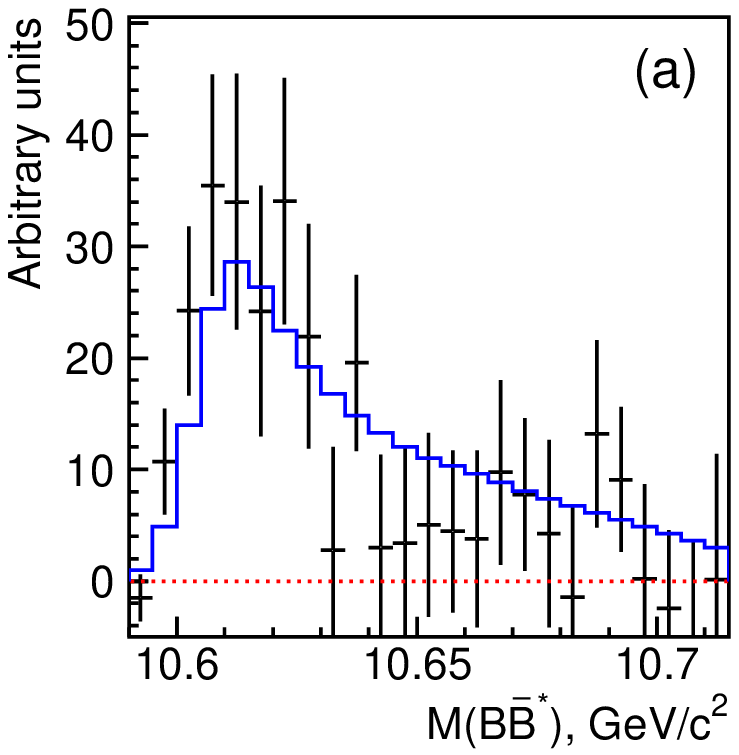,width=0.24\textwidth}
\epsfig{file=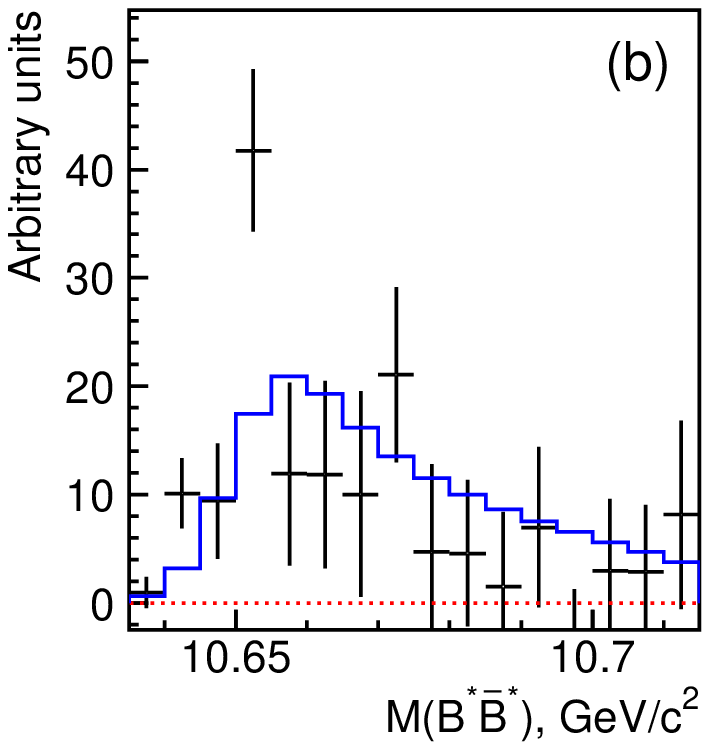,width=0.24\textwidth}
\epsfig{file=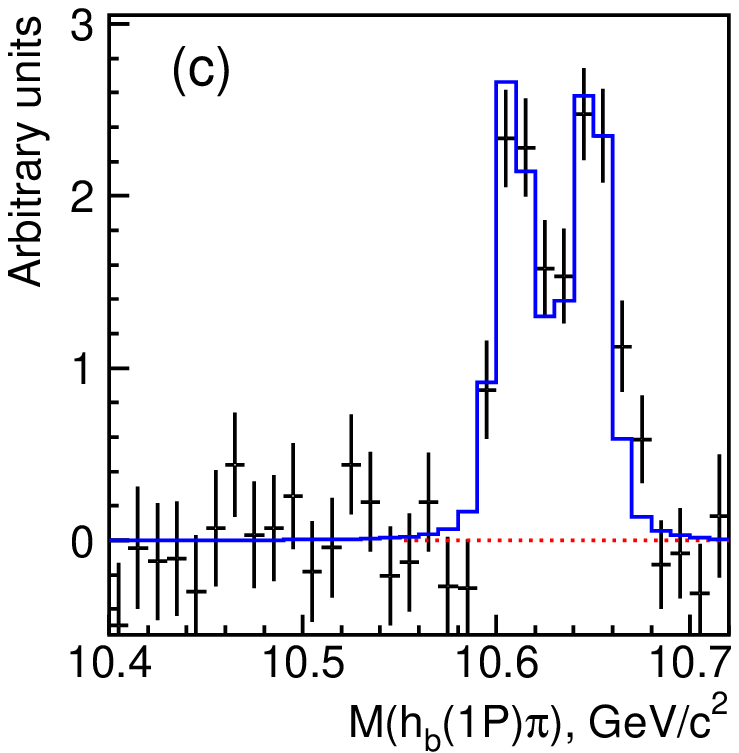,width=0.24\textwidth}
\epsfig{file=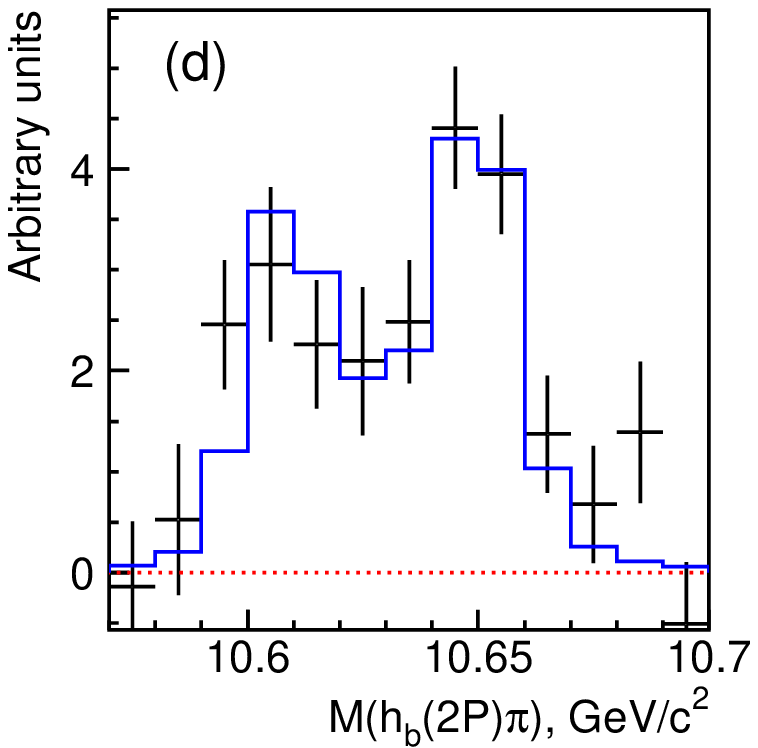,width=0.24\textwidth}\\
\epsfig{file=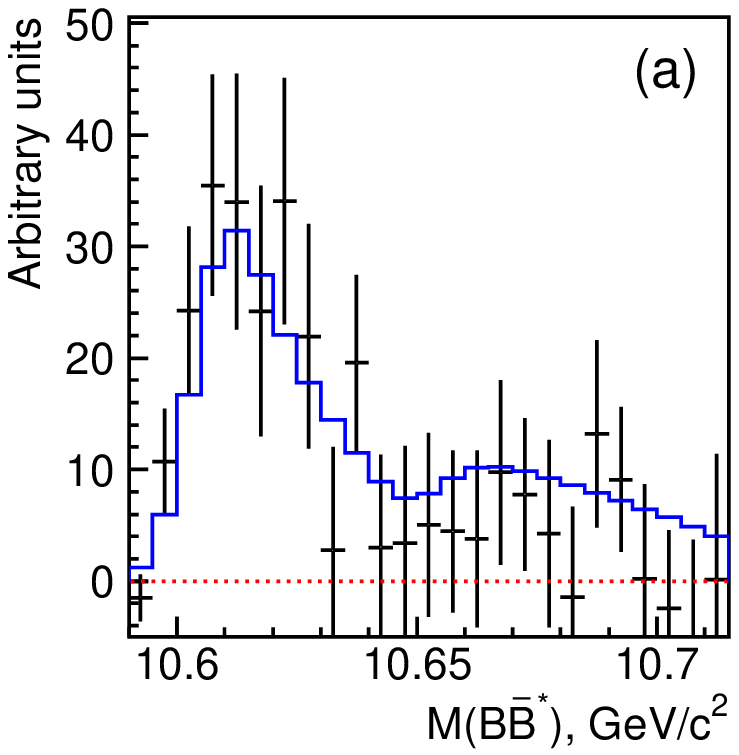,width=0.24\textwidth}
\epsfig{file=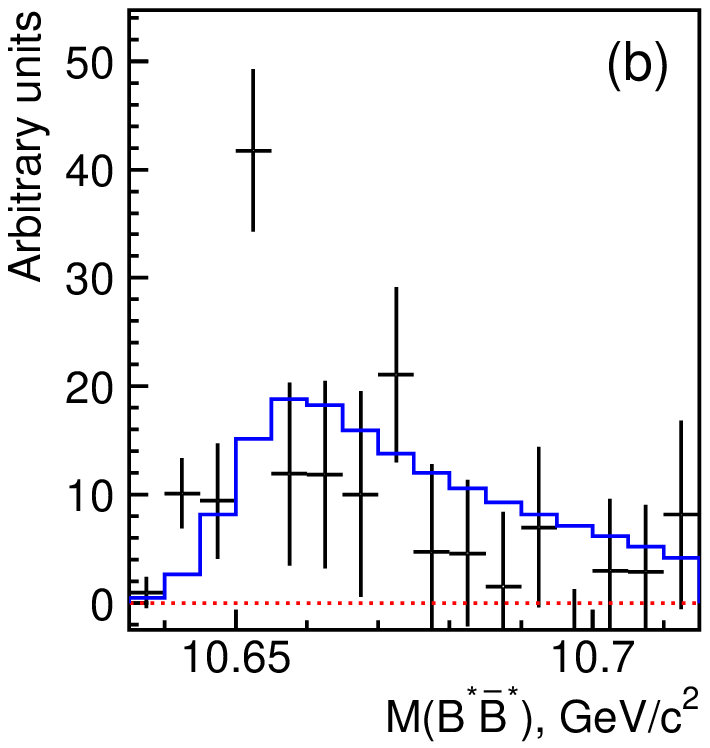,width=0.24\textwidth}
\epsfig{file=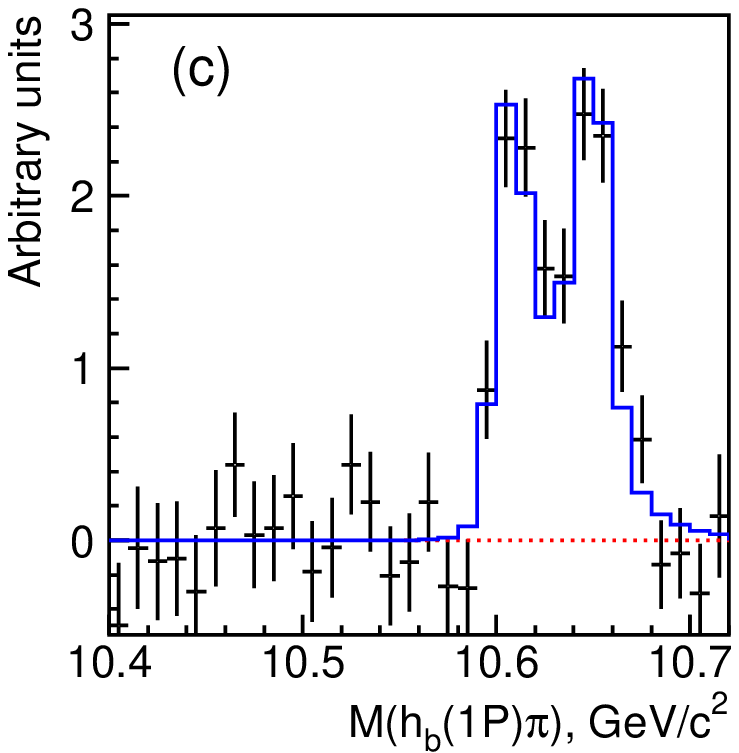,width=0.24\textwidth}
\epsfig{file=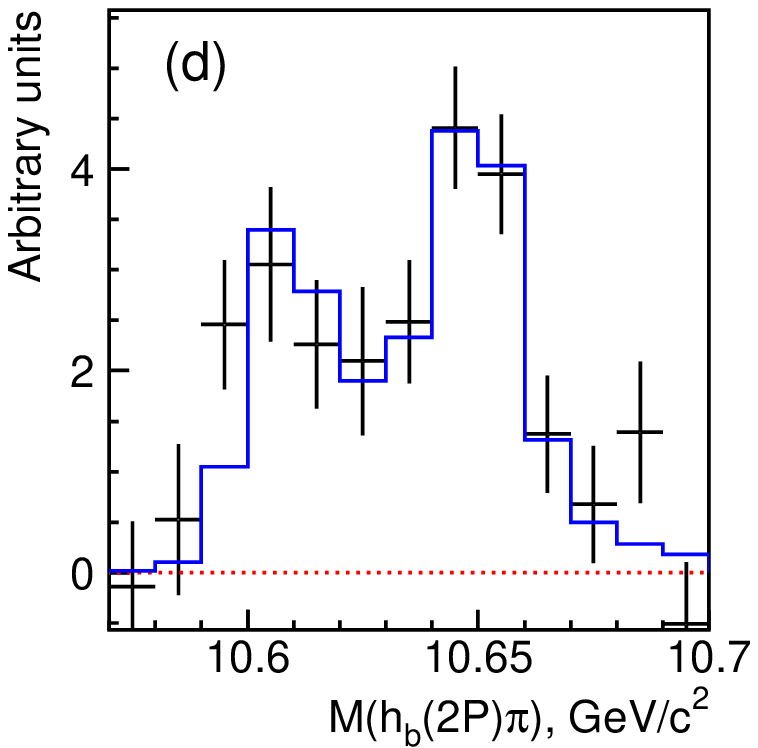,width=0.24\textwidth}
\end{center}
\caption{Fitted line shapes of the $Z_b(10610)$ and $Z_b(10650)$ in the $B^{(*)}\bar{B}^*$ channels [plots (a) and (b)]
and in the $\pi h_b(mP)$ ($m=1,2$) channels [plots (c) and (d)]. Parameters of fits A, B, and C are used for
the plots in the upper, middle, and lower rows, respectively.}\label{fig:fits}
\end{figure*}

We consider three different fits:

\noindent {\bf Fit A}. Combined fit for the data in the $Z_b^{(\prime)}\to \pi h_b(mP)$ ($m=1,2$) channels
\cite{Belle:2011aa} and
for the old data in the $Z_b^{(\prime)}\to B^{(*)}\bar{B}^{*}$ channels \cite{Adachi:2012cx} with HQSS constraints
(\ref{HSconstr}) and (\ref{HSconstrxi}) applied.
\smallskip

\noindent {\bf Fit B}. Same as fit A for the new data for the $Z_b^{(\prime)}\to B^{(*)}\bar{B}^{*}$ channels
\cite{Garmash:2015rfd}.
\smallskip

\noindent {\bf Fit C}. Same as fit B but with all parameters totally unconstrained.

The parameters of fits A, B, and C are quoted in Table~\ref{tab:params}, from which one can deduce several
conclusions. First, the suggested parametrisation is obviously able to capture all gross features of the experimental
signal and therefore provides a good overall description of the data in all
analysed channels. Second, one is led to conclude that the new data for the $Z_b^{(\prime)}\to
B^{(*)}\bar{B}^{*}$ channels are
much more compatible with the HQSS constraints. Indeed, on one hand, the quality of fit B is noticeably better
than the quality of fit A. Also, from fits B and C one can see that relaxing the HQSS
constraints does not lead to a considerable increase in the quality of the fit. This is to be confronted with the
dramatic decrease of the quality of the fit for the old data in the $Z_b^{(\prime)}\to B^{(*)}\bar{B}^{*}$
channels---from 76\%
for the totally unconstrained fit from Ref.~\cite{Hanhart:2015cua} to 32\% for fit A from Table~\ref{tab:params}.
Finally, fully unconstrained fit C demonstrates a better agreement with the HQSS constraints
(\ref{HSconstr}) and (\ref{HSconstrxi}) than the similar unconstrained fit to the old data found in
Ref.~\cite{Hanhart:2015cua}.

\begin{figure*}[t]
\begin{center}
\epsfig{file=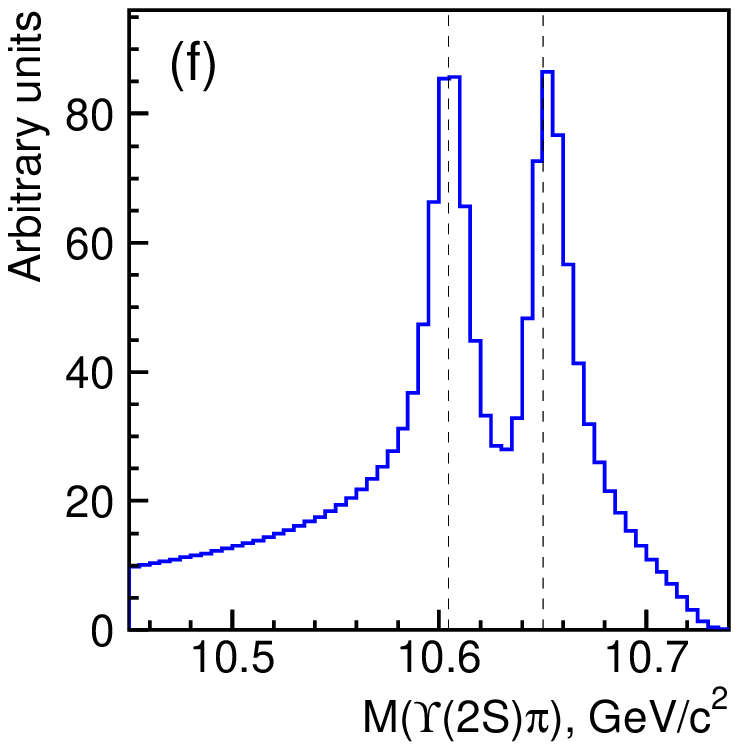,width=0.22\textwidth}
\epsfig{file=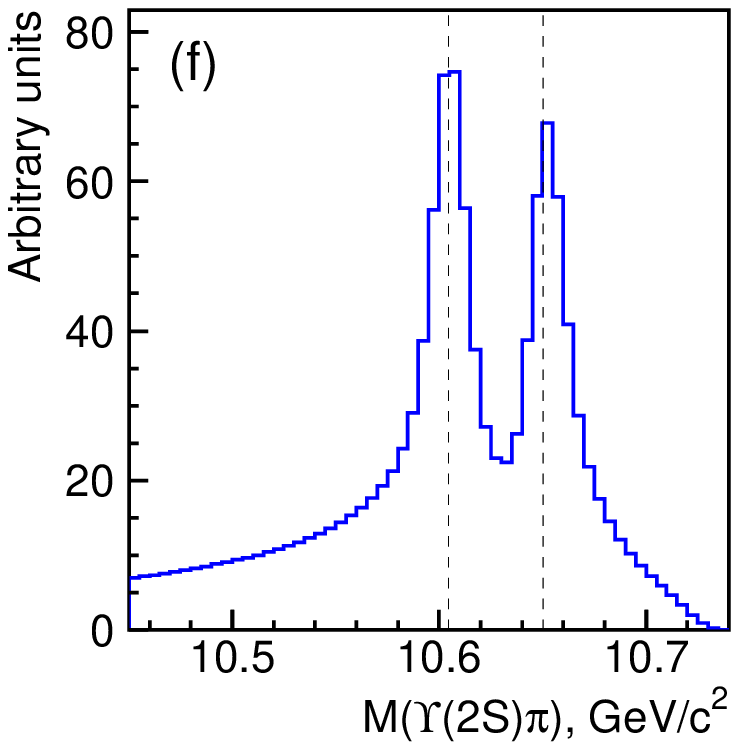,width=0.22\textwidth}
\epsfig{file=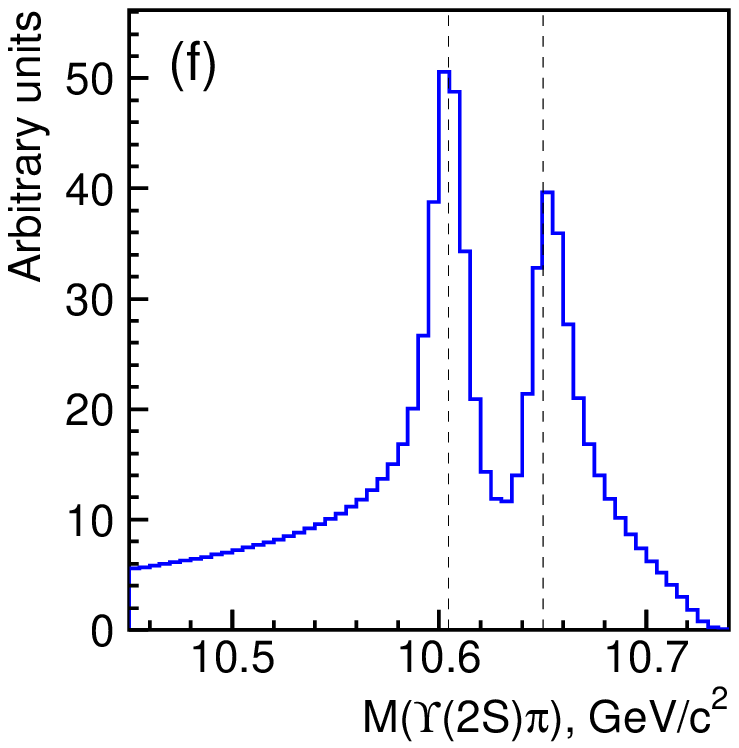,width=0.22\textwidth}
\epsfig{file=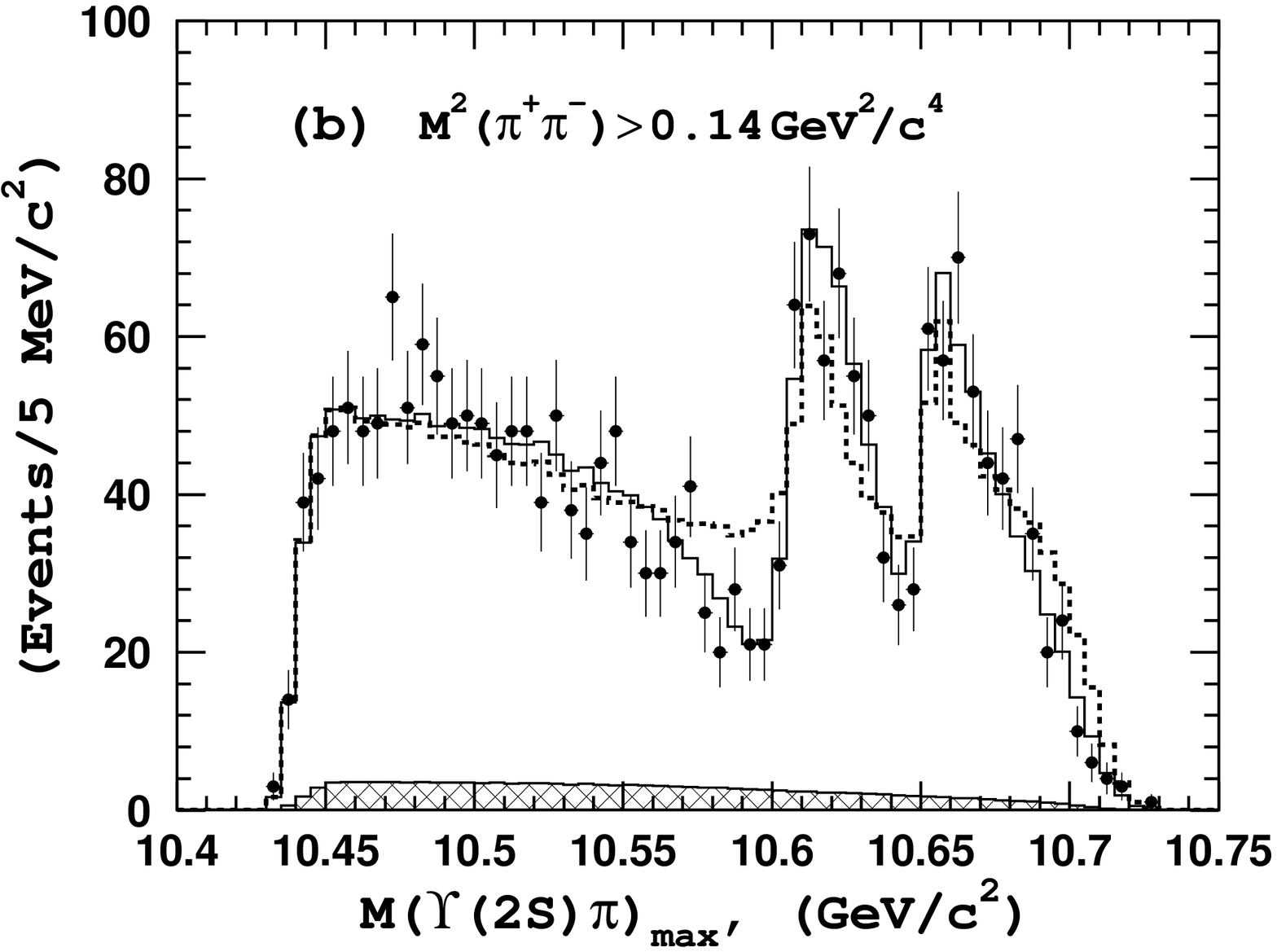,width=0.28\textwidth}
\end{center}
\caption{Predicted line shapes of the $Z_b(10610)$ and $Z_b(10650)$ in the $\pi\Upsilon(2S)$ channel for fits A, B, and
C, respectively. To guide the eye, as the last plot, we also show the corresponding experimental figure
adapted from Ref.~\cite{Garmash:2014dhx}. Notice that the behaviour of the line shape below the left shoulder of the
lower peak is influenced by the effects which lie beyond the scope of the present paper and
will be addressed in future publications. Notice also that the presence of the nonresonant background in the experimental figure
does not allow its direct comparison with the predicted line shapes.}\label{fig:fitsY}
\end{figure*}

The line shapes of the $Z_b(10610)$ and $Z_b(10650)$ states in the $B^{(*)}\bar{B}^*$ and $\pi h_b(mP)$ ($m=1,2$) channels are shown in 
Fig.~\ref{fig:fits} for all three fits from Table~\ref{tab:params}. In addition, as an example, we show, in Fig.~\ref{fig:fitsY}, the line shapes
in the $\pi\Upsilon(2S)$ channel which come as a prediction of our approach and demonstrate a clear similarity to the
experimental data (the last plot in Fig.~\ref{fig:fitsY}). The inclusion of the
information on the $\pi\Upsilon(nS)$ line shapes in a future multidimensional
analysis will help to improve the accuracy of the determination of the model
parameters.

Two comments on the fits given in Table~\ref{tab:params} are in order
here:

(1) While fits A and B have the HQSS constraints built in, fit C features some
HQSS breaking since $\xi$ takes a value different from $-1$ (see
Eq.~(\ref{HSconstrxi})) and, particularly, since the ratios $g_{[\pi
h_b(nP)][B^*\bar{B}^*]}/g_{[\pi h_b(nP)][B\bar{B}^*]}~(n=1,2)$ deviate from
their respective HQSS values (\ref{HSconstr}).
This might be because of the complexity of the $\Upsilon(10860)$ state, assigned
as the $5S$ bottomonium here, so that the HQSS breaking effects may stem from a
mixture of the $D$-wave bottomonium~\cite{Guo:2014qra} or non-$\bar{b}b$
components~\cite{Ali:2009es} in the $\Upsilon(10860)$ wave function.
It is worthwhile noticing that, even in the two-body open-bottom decays of the
$\Upsilon(5S)$, the measured branching fractions~\cite{Agashe:2014kda} show a sizable HQSS
breaking as well. This was summarised, for example, in
Ref.~\cite{Drutskoy:2012gt}. It is also concluded in Ref.~\cite{Mehen:2013mva}
that explicit HQSS breaking operators are needed to describe the $Z_b$'s line
shapes in the $\Upsilon(5S)\to \pi B^{(*)}\bar B^*$ decays. On the other hand, this
deviation may be diminished in the fit to updated experimental data in the
future. If, however, the HQSS breaking still persists, one will need to
investigate the origin carefully since HQSS is normally very well respected in
the bottomonium mass region. In addition to the possible non-$S$-wave $\bar{b}b$
component for the $\Upsilon(10860)$, the internal dynamics of the $Z_b$ states
might be another reason. However, this breaking seems to be rather unlikely to
occur due to the reason discussed in Ref.~\cite{Cleven:2011gp} where a large
HQSS breaking effect in the ratio $g_{Z_b B\bar B^*}/g_{Z_b B^*\bar B^*}$ is
explained by the proximity of the poles to the corresponding thresholds. Such an
effect manifests itself in the pole positions of the amplitude and therefore it
was already included in the fits. Furthermore, it was pointed out in
Ref.~\cite{Voloshin:2013ez} that the $S$-$D$ mixing effects for the bottom meson
pair in the final state of the decay $\Upsilon(5S)\to \pi B^{(*)}\bar B^*$
probably only play a minor role for the internal structure of the $Z_b$ states
(see also Ref.~\cite{Sun:2011uh} for a calculation based on the one-meson
exchange model).

(2) In fit B which has HQSS built in, the values of $\gamma_s$ and $\gamma_t$ are
almost the same. It means that the off-diagonal matrix elements of the potential
matrix for the interaction between elastic channels almost vanish. Indeed, from
Eqs.~(\ref{vs}) and (\ref{gammast}) and for the parameters of fit B, we have
\be
v_{12}=\frac{1}{8\pi^2\mu}\left(\gamma_s^{-1}-\gamma_t^{-1}\right)\ll v_{11}=v_{22}.
\label{eq:lqss}
\ee
Since $\gamma_s^{-1}\propto \VII$ and $\gamma_t^{-1}\propto \VI$ describe the interaction for the total light-quark spin 1 and 0, respectively
[see Eq.~(\ref{VVV})], this is in fact consistent with the observation made recently~\cite{Voloshin:2016cgm}
that the nonobservation of the $Z_b(10650)$ in
the $B\bar B^*$ invariant mass distribution implies that the interaction between the
bottom and antibottom mesons is insensitive to the
light quark spin, and thus seems to imply an accidental ``light-quark spin symmetry.''
Indeed, there is little signal of the $Z_b(10650)$ in  the plot (a) in the
second row of Fig.~\ref{fig:fits}. However, although not prominent, the
$Z_b(10650)$ shows up as a bump in the plot (a) in the third row of
Fig.~\ref{fig:fits}, which corresponds to fit~C with HQSS constraints released.
In this fit, $\gamma_s$ and $\gamma_t$ do not take similar values any more. This
means that the current data require us to understand either the accidental
light-quark spin symmetry or a sizable HQSS breaking.

For completeness, we quote all parameters of fit B in Table~\ref{tab:paramsfitB}.

\begin{table*}
\begin{ruledtabular}
\begin{tabular}{cccccc}
Fit &$g_{[\pi h_b(1P)][B\bar{B}^*]}\cdot 10^3$&$g_{[\pi h_b(2P)][B\bar{B}^*]}\cdot 10^3$&
$g_{[\pi \Upsilon(1S)][B\bar{B}^*]}\cdot 10^4$&$g_{[\pi \Upsilon(2S)][B\bar{B}^*]}\cdot 10^4$&$g_{[\pi \Upsilon(3S)][B\bar{B}^*]}\cdot 10^4$\\
\hline
B &
$2.0^{+0.3}_{-0.2}$&
$7.5^{+1.0}_{-0.9}$&
$1.3  \pm 0.3$&
$5.0^{+0.8}_{-0.9}$&
$7.0^{+1.3}_{-1.5}$\\
\hline
C &
$1.2^{+0.5}_{-0.4}$&
$4.6^{+1.7}_{-1.4}$&
$1.4  \pm 0.3$&
$5.5  \pm 1.0$&
$7.9^{+1.6}_{-1.8}$
\end{tabular}
\end{ruledtabular}
\caption{Parameters of fits~B and C. The couplings $g_{[\pi h_b(mP)][B\bar{B}^*]}$ and $g_{[\pi \Upsilon(nS)][B\bar{B}^*]}$ are 
given in the units of GeV$^{-3}$ and GeV$^{-2}$, respectively. For both fits,  
$g_{[\pi\Upsilon(nS)][B^*\bar{B}^*]}/g_{[\pi\Upsilon(nS)][B\bar{B}^*]}=-1$, as required by the HQSS constraints from Eq.(\ref{HSconstr}), while 
the values of the ratios $g_{[\pi h_b(mP)][B^*\bar{B}^*]}/g_{[\pi h_b(mP)][B\bar{B}^*]}$ can be found in 
Table~\ref{tab:params}.}\label{tab:paramsfitB}
\end{table*}

\section{Nature of the $Z_b(10610)$ and $Z_b(10650)$ from data}
\label{sec:NatureZb}

Important information on the nature of the near-threshold states like the $Z_b(10610)$ and $Z_b(10650)$ is encoded in the singularity structure
of the amplitudes extracted from the fit,\footnote{It has to be noticed that the obtained values of the parameters cannot be compared directly with
those from, e.g., Ref.~\cite{Nieves:2012tt} since, in the latter paper, a Gaussian vertex form
factor was used to regularise the Lippmann-Schwinger equation and the contact
terms are scale-dependent.} in particular the pole positions and pole residues~\cite{Weinberg:1962hj,Baru:2003qq,Aceti:2012dd,Nagahiro:2014mba}.
Therefore we have a closer look at the pole locations of the $Z_b$ states in this section.

The full $t$ matrix considered here has in total seven coupled channels. One might
think that the task of searching for the poles of the $t$ matrix is formidable,
because the number of Riemann sheets is $2^7=128$. However, in practice the
problem is as simple as a two-channel one. This is because the thresholds of all
the inelastic channels are far away from those of the $B\bar B^*$ and $B^*\bar
B^*$ channels and the interactions among the inelastic channels are very weak and can be safely
neglected as it is anyhow done in this paper. Thus any pole which has the
potential to produce a measurable effect should reside well above all the
inelastic thresholds. Therefore, the relevant Riemann-sheet structure is
practically the same as that for the two-channel case.

In order to search for the poles in these relevant Riemann sheets, one needs to put
all the inelastic channels in their corresponding unphysical sheets.
This is achieved by an analytic continuation with a practical trick of changing
the sign of the imaginary part of the inelastic channel Green's functions given
in Eqs.~(\ref{eq:G0in}), (\ref{G0}) and (\ref{eq:Valphaa}).

\begin{figure*}
\begin{center}
\centerline{\includegraphics[width=0.28\linewidth]{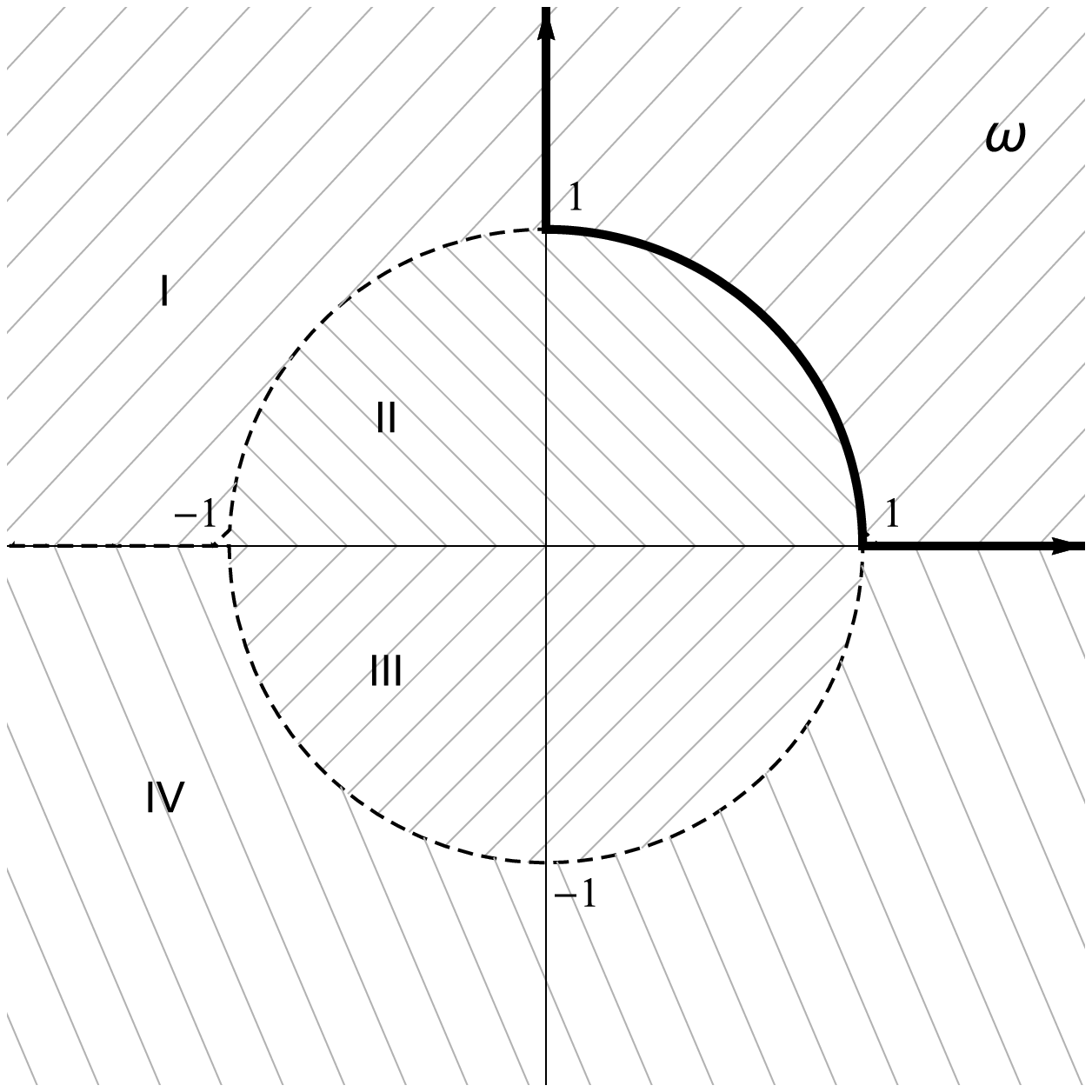}\hspace*{0.02\textwidth}
\includegraphics[width=0.3\textwidth]{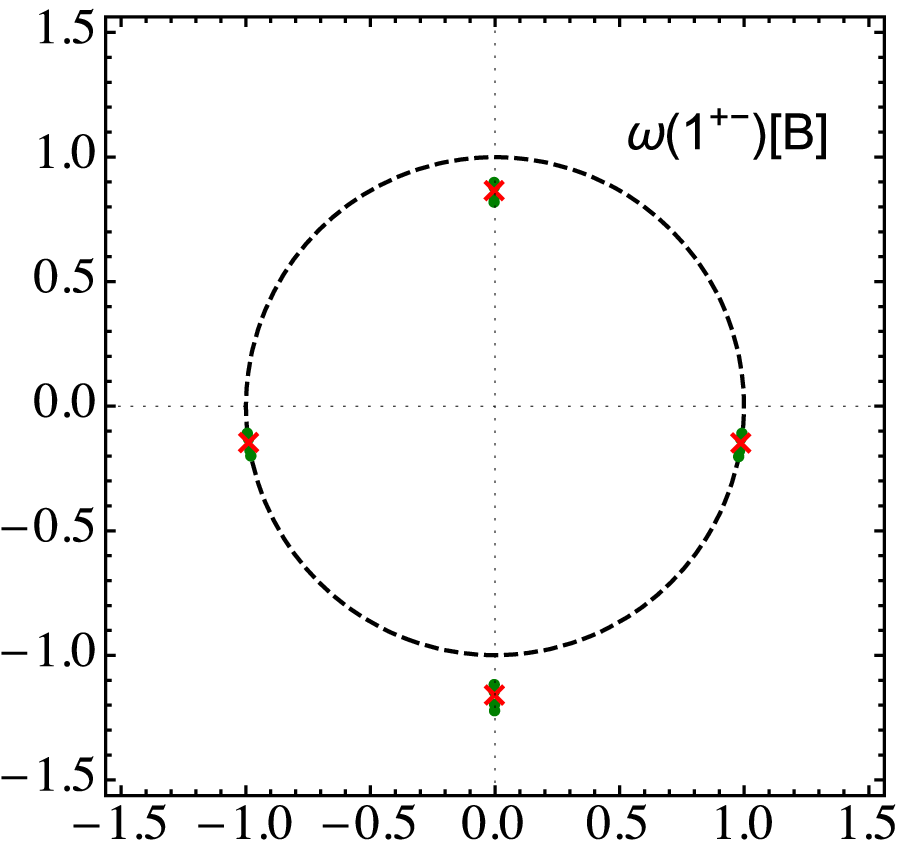}\hspace*{0.02\textwidth}
\includegraphics[width=0.3\textwidth]{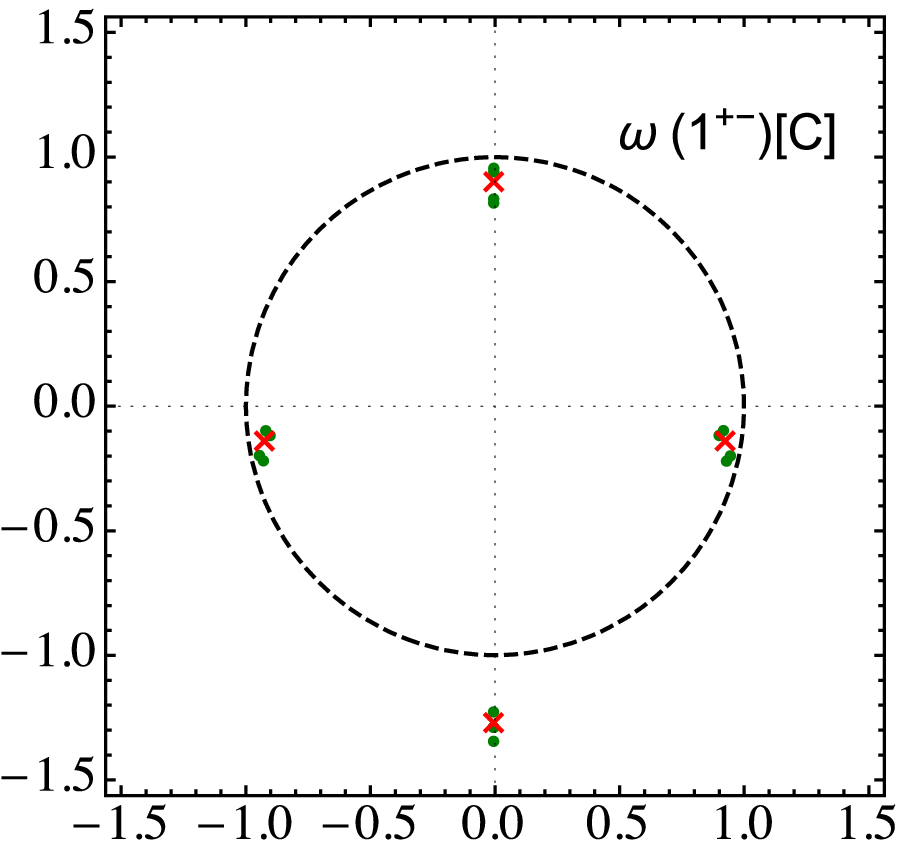}}
\caption{First plot: Four Riemann sheets mapped into the $\omega$ plane. The thick solid line corresponds to the real axis in the first Riemann sheet
of the complex energy plane. Second plot: The poles of the full $t$ matrix which correspond to the quantum numbers $1^{+-}$ and to the set of
parameters from fit B---see Table~\ref{tab:params}. The red crosses mark the central values and the green dots show the upper and
the lower bounds due to the uncertainties in the fitted parameters. Third plot: The same as in the second plot but for the set
of parameters from fit C.}
\label{omegaFig-fitB}
\end{center}
\end{figure*}

To study the poles in the two-channel case with the quantum numbers $1^{+-}$, it is convenient to make a
conformal mapping from the four-Riemann-sheet complex energy plane to the single complex $\omega$ plane~\cite{kato}.
For a given energy $E$, we can write
\be
E=\frac{k_1^2}{2\mu}=\frac{k_2^2}{2\mu}+\delta,
\label{Ek1k2}
\ee
where $\delta=m_{B^*}-m_B$ denotes the energy gap between the two elastic thresholds. Instead of two complex momenta $k_1$
and $k_2$ constrained by the two conditions from Eq.~(\ref{Ek1k2}), we switch to the complex variable $\omega$, defined via
\be
k_1=\sqrt{\frac{\mu\delta}{2}}\left(\omega+\frac{1}{\omega}\right),\quad
k_2=\sqrt{\frac{\mu\delta}{2}}\left(\omega-\frac{1}{\omega}\right).
\label{omegak1k2}
\ee
This allows us to rewrite the energy as
\be
E=\frac{\delta}{4}\left(\omega^2+\frac{1}{\omega^2}+2\right).
\ee
By construction, the complex $\omega$ plane is free of unitary cuts.

In the first plot in Fig.~\ref{omegaFig-fitB} we show the mapping of the four Riemann sheets of the complex energy plane, labelled as
\begin{eqnarray}
\begin{aligned}
\text{RS-I:}&&\quad{\rm Im}~k_1>0,\quad{\rm Im}~k_2>0,\\
\text{RS-II:}&&\quad{\rm Im}~k_1<0,\quad{\rm Im}~k_2>0,\\
\text{RS-III:}&&\quad{\rm Im}~k_1>0,\quad{\rm Im}~k_2<0,\\
\text{RS-IV:}&&\quad{\rm Im}~k_1<0,\quad{\rm Im}~k_2<0,
\label{eq-RS}
\end{aligned}
\end{eqnarray}
onto the $\omega$ complex plane. The thick solid line corresponds to real values of the energy $E$ on the first sheet,
and the part of the imaginary $\omega$ axis with $\text{Im}\,\omega>1$ corresponds to negative values of $E$, thus
representing energies below the $B\bar B^*$ threshold.

It is easy to see from Eq.~(\ref{omegak1k2}) that the $B\bar B^*$ threshold ($k_1=0$) appears at $\omega=\pm i$ and the $B^*\bar
B^*$  ($k_2=0$) threshold appears at $\omega=\pm1$. Thus the near-threshold regions correspond to the vicinities of $|\omega|=1$.
To be able to distinguish between the poles according to their relevance for producing structures in the amplitude in
the physical region, it is worthwhile to discuss the structure of the Riemann sheets in some more detail.
In particular, between the thresholds, RS-I is glued with RS-II and RS-III is glued with RS-IV along the real energy
axis, since crossing this axis changes the sign of Im\,$k_1$. Above the higher threshold, crossing the real energy
axis changes the signs of both Im\,$k_1$ and Im\,$k_2$ so that, in this region, RS-I is attached to RS-IV and
RS-II is attached to RS-III.

We find that the $1^{+-}$ $t$ matrix possesses four poles in the
complex $\omega$ plane, shown in Fig.~\ref{omegaFig-fitB}. The pole near the
imaginary axis in the lower half $\omega$ plane corresponds to a pole below
the $B\bar B^*$ threshold lying on RS-IV of the complex energy plane.
It therefore appears far away from the physical region and has little impact on the
physical amplitude. It will not be discussed below.

The pole in the upper half $\omega$ plane (if we switch off the inelastic channels, it is located exactly on the
imaginary axis) lies nearly on the real axis on RS-II of
the complex energy plane, so it describes a virtual state. It is close to the
$B\bar B^*$ threshold and corresponds to the $Z_{b}(10610)$.
The nonzero real part of the pole location in the $\omega$-plane (which translates into
a finite imaginary part in the energy plane) reflects the fact that the $Z_b(10610)$ can decay into the inelastic channels. Notice that, for the
parameters
from fit C, $v_{11}\propto \gamma_s^{-1}+\gamma_t^{-1}>0$ and therefore, in the single-channel case (neglecting the $B^*\bar B^*$ channel), the $t$
matrix
\be
t\propto\frac{1}{v_{11}^{-1} +i\, (2\pi)^2\mu k_1}
\ee
would have a bound-state pole. However, in the two-channel case, the pole in the vicinity of the $B\bar B^*$ threshold
is a virtual state. This means the $B^*\bar B^*$ channel effectively reduces the attraction in the $B\bar B^*$
system and turns the bound state into a virtual state. For the parameters from fit B the $Z_b(10610)$ pole corresponds to a virtual state both in
the single-channel and two-channel case.

\begin{figure}
\centering
\includegraphics[width=\linewidth]{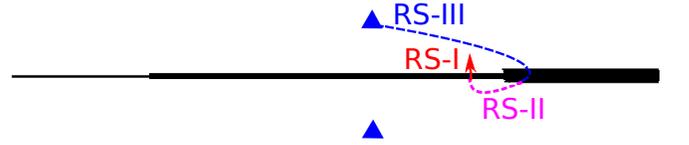}
\caption{The path of the RS-III pole to RS-I. The complex conjugated pole is
also shown but not its path.}
\label{fig:rsiii}
\end{figure}

The other two poles, with $\omega\simeq\pm 1$, are a pair of conjugated poles
below the $B^*\bar B^*$ threshold. We focus on the right one, for it is this
pole that is closest to the physical region. This pole lies on RS-IV (RS-III)
for fit B (C) and corresponds to the $Z_b(10650)$.
The nonzero imaginary part of the pole reflects the fact that $Z_b(10650)$ can
decay into the lower $B\bar B^*$ channel as well as into the inelastic channels.
This pole is very close to the $B^*\bar B^*$ threshold and as such it is able to
produce a pronounced peak in the line shape. For fit C, the path from
the pole in RS-III to the physical RS-I is to go up to the $B^*\bar B^*$
threshold, to enter RS-II and then to approach RS-I from below the $B^*\bar B^*$
threshold---see the sketch in Fig.~\ref{fig:rsiii} (or to go to RS-IV from
below the $B^*\bar B^*$ threshold and then approach RS-I from above that
threshold).
For fit B the pole appears on RS-IV and therefore it has a simpler path to the physical region by crossing the cut
above the $B^*\bar B^*$ threshold since RS-I and RS-IV are directly glued there.

The $Z_b$ and $Z_b'$ energies relative to the respective thresholds,
\bea
\varepsilon_B(Z_b)&\equiv& M(B\bar{B}^*)-M(Z_b), \label{EBZbs}\nonumber\\[-2mm]
\label{eq:BindingEnergy}\\[-2mm]
\varepsilon_B(Z_b')&\equiv& M(B^*\bar{B}^*)-M(Z_b'),\nonumber
\eea
are
\bea
\varepsilon_B(Z_b)&=&(1.10_{-0.54}^{+0.79}\pm i 0.06_{-0.02}^{+0.02})~\mbox{MeV},\nonumber\\[-1mm]
\label{EBZbs1}\\[-1mm]
\varepsilon_B(Z_b')&=&(1.10^{+0.79}_{-0.53} \pm i 0.08_{-0.05}^{+0.03})~\mbox{MeV},\nonumber
\eea
for the parameters from fit B, and
\bea
\varepsilon_B(Z_b)&=&(0.60_{-0.49}^{+1.40}\pm i 0.02_{-0.01}^{+0.02})~\mbox{MeV},\nonumber\\[-1mm]
\label{EBZbs2}\\[-1mm]
\varepsilon_B(Z_b')&=&(0.97^{+1.42}_{-0.68} \pm i 0.84_{-0.34}^{+0.22})~\mbox{MeV},\nonumber
\eea
for the parameters from fit C. In order to determine the uncertainties of the pole positions we varied the parameters
$\gamma_s$ and $\gamma_t$ within their ranges allowed by the respective fit. We
notice that the real parts of the poles are always below the corresponding
thresholds. In addition, the close similarity of the two pole positions for fit B
is again a consequence of nearly vanishing $v_{12}$---see the discussion around
Eq.~\eqref{eq:lqss}.

As one can see, the current data are consistent with both $Z_b(10610)$ and $Z_b(10650)$ as virtual states.
This may have severe implications for the interpretation of their nature, since only states with a dominant
two-hadron component can be virtual states.\footnote{By solving the Schr\"odinger equation for a four-quark system, tetraquark states correspond to
the bound states of four quarks and thus cannot be virtual states.} Thus our findings give a strong support to the conjecture that the two $Z_b$
states qualify as hadronic molecules. Meanwhile, improved data are necessary to confirm this conclusion.

\begin{figure}[tbh]
\centering
\includegraphics[width=0.4\textwidth]{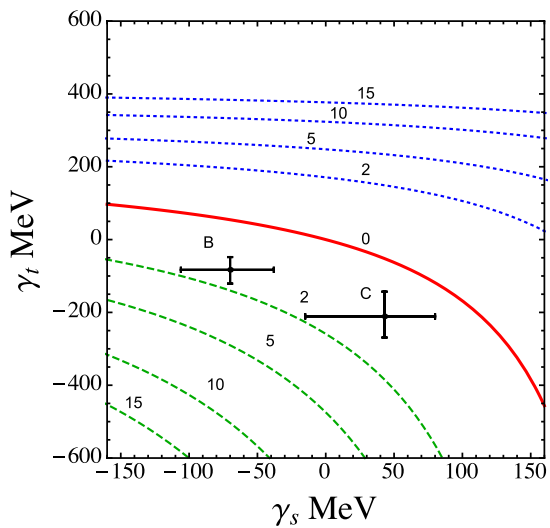}
\caption{The space of parameters ($\gamma_t$ versus $\gamma_s$) for the poles of $t^v$ in the channel $1^{+-}$ close to the $B\bar B^*$ threshold
[$Z_b(10610)$ state]. The blue and green curves correspond to different Riemann sheets (RS-I versus RS-II, respectively) and the red line gives the
boundary
between the two regions. The energy of the state relative to the threshold is quoted, in MeV, at every curve. The black dots with the error bars
show the actual values of the parameters $\gamma_s$ and $\gamma_t$ for fits B and C taken from Table~\ref{tab:params}.}
\label{fig-parameterBE1}
\end{figure}

In the remainder of this section we demonstrate how well the pole locations are determined by the data currently available.
To proceed in this direction we stick to the $Z_b(10610)$ pole and consider fits B and C. We observe that the parameters of the fits do not
change appreciably, if only the direct interaction $t$ matrix $\tv$ is retained in the elastic $t$ matrix.
We also notice that, in the current data set, the influence of the
inelastic channels on the line shapes is not very strong either,
their role being mainly to provide a finite imaginary part to the poles. We
therefore now study the poles of just the direct interaction $t$ matrix, $t^v$,
which depends only on $\gamma_s$ and $\gamma_t$. In the
$(\gamma_s,\gamma_t)$-plane we identify various regions, which correspond
to different Riemann sheets---see Fig.~\ref{fig-parameterBE1}. The actual values of the parameters $\gamma_s$ and $\gamma_t$
taken from fits B and C are shown by the black dots with the error bars. The red curve
\begin{eqnarray}
\gamma_t=\left(\gamma_s^{-1}-\sqrt{2/(\mu\delta)}\right)^{-1}
\end{eqnarray}
separates the parameter space for the $Z_b(10610)$ as a virtual state from that for the $Z_b(10610)$ as a
bound state. Then each blue (green) curve corresponds to a bound (virtual) state with the pole energy, relative to
the $B\bar B^*$ threshold, quoted explicitly, in MeV, near every curve.

From Fig.~\ref{fig-parameterBE1} one can see that, while the data are rather
uncertain and the parameters $\gamma_s$ and $\gamma_t$ found from different fits
differ substantially, the corresponding dots in the $(\gamma_s,\gamma_t)$-plane
nevertheless reside in the ``green'' domain (virtual state) sufficiently far
away from the red boundary curve. Therefore, the conclusion that the $Z_b(10610)$ is
a virtual state can be treated as a robust prediction from the data. A similar
conclusion holds concerning the nature of the $Z_b(10650)$ as a resonance;
however, even in the absence of the inelastic channels, the $Z_b(10650)$ pole
has an imaginary part and therefore its fate cannot be demonstrated in a plot as
simple as that for the $Z_b(10610)$ given in Fig.~\ref{fig-parameterBE1}.

\section{Remarks on the possible role of pion exchanges}

The role of one-pion exchange (OPE) on the formation of exotic resonances and, in particular,
of hadronic molecules is discussed heavily in the literature.
While Refs.~\cite{Fleming:2007rp,Nieves:2011vw,Nieves:2012tt} argue that this contribution to the potential is
perturbative, Refs.~\cite{Tornqvist:1991ks,Swanson:2003tb} claim it to be a crucial contribution
to the binding of the two-hadron system.

It was stressed in Ref.~\cite{Baru:2015nea} that from the point of view of field theoretical consistency
the significance of the OPE for the binding energy of the charmonium state $X(3872)$ cannot be defined unambiguously. Given an apparent similarity of
the pion exchanges between $D^{(*)}$ mesons and $B^{(*)}$ mesons, the same conclusion holds for the $Z_b$'s.
Meanwhile, the long-range tail of the OPE potential might distort the $Z_b$'s line shapes significantly~\cite{Voloshin:2015ypa}.
In addition, it might also induce a significant mixing between the $B\bar B^*$ and $B^*\bar B^*$
channels as observed in Ref.~\cite{Voloshin:2016cgm}. We therefore briefly comment on the
possible role of the OPE here---a detailed calculation including pion exchanges will
be presented in a subsequent publication~\cite{prep}.

Clearly, the leading effects that determine the line shapes are the pole positions of the two
$Z_b$ resonances. In the analysis of the existing data presented above the pole locations emerged
from a subtle interplay of the channel couplings. We expect this pattern to persist also when
pion exchanges are included, since still free parameters can be adjusted to locate the poles to
where data request them to be. Effectively this means that, compared to this analysis, the
pion exchange can at most slightly vary the $Z_b$ line shapes. In particular, we do not expect this effect
to be as large as announced in Ref.~\cite{Voloshin:2015ypa} for two reasons: first of all,
the analysis of this work did not consider the effect of the interplay of the two poles (determined
by their location in the complex plane) on the experimental signals and, secondly, the
effect of the OPE was maximised in Ref.~\cite{Voloshin:2015ypa} by using an
effective pion mass $\mu_\pi = \sqrt{m_\pi^2-\delta^2}$, with
$\delta=m_{B^*}-m_B$, in the expression for the static OPE.
However, this kind of OPE is correct only
for the on-shell potential
in the $B\bar B^*$ channel and takes a different structure
in the $B^*\bar B^*$ channel as well as for the transition potential.
In addition, the (half-)off-shell potential, relevant here, is
energy dependent and
when spanning an energy range that covers both $Z_b$ states and
keeping effects of the order of $\delta$, also the energy dependence of the
OPE potential needs to be kept, which is of the same order. This changes the effective pion
mass in a nontrivial way over the relevant energy range.
It is also important to keep in mind that as soon as the energy dependence of the
pion exchange contribution is to be kept, the recoil terms of the $B$ mesons
need to be kept as well, for they contribute to the same order, as stressed in Ref.~\cite{Hanhart:2007mu}
in a different context.
Similar arguments as the ones just presented also allow one to question the
claim of Ref.~\cite{Voloshin:2016cgm} that the contribution of the OPE spoils the light quark spin symmetry.
More details will be given elsewhere~\cite{prep}.

Therefore, to summarise
the arguments just presented, we expect that even if OPE were included
in the analysis of the data for the $Z_b$ states the line shapes would change only slightly.
It should be stressed that regardless of this claim a systematic study of the pion exchange
contribution to exotic states is still very valuable. For example, the quark mass dependence of exotic states can only be
studied in a controlled way with this contribution included \cite{Baru:2013rta,Jansen:2013cba,Baru:2015tfa}.
This is of relevance for chiral extrapolations of lattice data that at present
exist only at unphysically high quark masses~\cite{Prelovsek:2013cra}.
Another example of the relevance of the OPE for studies of
exotic states is given in Ref.~\cite{Cleven:2015era}, where  it is pointed out that
it leads to a very specific pattern of exotic states with respect to their quantum numbers.

\section{Summary}\label{sum}

In this paper we formulate and analytically solve a co\-upl\-ed-channel problem
for the scattering $t$ matrix involving elementary states and a set of elastic
and inelastic channels coupled to each other. The solution found can be viewed
as a further generalisation of the approach presented before in
Refs.~\cite{Baru:2010ww,Hanhart:2011jz}. It should be stressed that since the
approach is based on the Lippmann-Schwinger equations for the coupled-channel
problem, all unitarity and analyticity constraints for the $t$ matrix are
fulfilled automatically. In particular, in contrast to earlier works, the
inelastic channels are taken into account non\-per\-tur\-ba\-ti\-ve\-ly; that is
they are iterated to all orders. Then unitarity guarantees that all imaginary
parts are included in a selfconsistent way. On the other hand, since to leading
order in a low-energy expansion there is no direct interaction within the
inelastic channels, at least for the type of the systems discussed here, the
inelastic channels enter the expressions only additively.
As a result it is very easy to include additional inelastic channels.

We present a parametrisation of the solution of the equations which appears to
be relatively simple but should be powerful enough to describe line shapes of
near-thre\-shold states in a wide class of reactions. As a byproduct of the
explicit unitarity of the approach, the suggested parametrisation allows one
to test the existing experimental data for completeness. Indeed, if there exist
not yet measured inelastic channels coupled to the elastic ones, the former will
contribute to the inelasticities (\ref{Gcalin}) and (\ref{G0}).
The corresponding contributions would induce additional imaginary parts of the
effective potentials, not linked to the decays known experimentally. If the best
fit to all existing data gives negligibly small values of these additional
inelasticities, the model can be regarded as complete up to the precision of
the experimental data.
On the contrary, large values of the additional imaginary parts would indicate a large violation of
unitarity which can only be recovered by enlarging the basis of the channels
explicitly included in the model. This would also mean that additional
experimental efforts are necessary to identify and to measure the missing
inelastic channels.

Finally, we exemplify the suggested approach by the line shapes for the
bottomoniumlike states $Z_b$ and $Z_b'$.
Without introducing any elementary state, the experimental data for the $Z_b$
and $Z_b'$ can be described well, and poles corresponding to these two states
with $J^{PC}=1^{+-}$ and $I=1$ are found in the $t$ matrix. We conclude that the $Z_b(10610)$ is a virtual state
located on the second Riemann sheet near the $B\bar{B}^*$ threshold while the $Z_b(10650)$ is a resonance on the third or fourth
Riemann sheet (however very close to the first Riemann sheet) lying near the $B^*\bar{B}^*$ threshold.

With the parameters extracted from the combined fit for the data, pole positions can be predicted in the
complementary channels, with the quantum numbers $0^{++}$, $1^{++}$, and $2^{++}$, in addition to those which
have the quantum numbers $1^{+-}$ and correspond to the $Z_b(10610)$ and $Z_b(10650)$ states.
The presence of such additional isovector poles complies very well with the expectations of the existence of
more isovector hidden-bottom hadronic molecules, called $W_b$---see Refs.~\cite{Voloshin:2011qa,Mehen:2011yh}.
However we refrain from further dwelling on the $W_b$'s here because their study requires some caution and, in
particular, might call for the inclusion of the pion exchanges. We therefore leave this for
future publications.

Unfortunately, with the present quality of the data,
the parameters extracted from the fits are very uncertain (notice, for example, the opposite signs of the parameter
$\gamma_s$ in fits B and C as well as a factor 3 difference in $\gamma_t$, while both fits provide a similar good overall
description of the data) and so are the predictions for the pole positions found with the help of these parameters. It
is expected however that future high statistics and high resolution experiments should provide more accurate and more
complete data sets.

Finally, we argue that the contribution of the nonseparable one-pion exchange potential is small, once the parameters
are refitted with pion exchanges included. As a result, it
should be safe to apply the parametrisation scheme presented here also to further experimental analyses.
In particular, the use of sums of Breit-Wigner functions should be abandoned for the analysis of near-threshold states.

\begin{acknowledgments}
We would like to thank Alexander Bondar, Martin Cleven, Johann Haidenbauer and Andreas Nogga for valuable discussions.
This work is supported in part by the DFG and the NSFC through funds provided to the Sino-German CRC 110 ``Symmetries and the Emergence of Structure
in QCD'' (NSFC Grant No. 11261130311). R. M. and A. N. acknowledge support from the Russian Science Foundation (Grant No. 15-12-30014). F.-K.~G. is
partially supported by the Thousand Talents Plan for Young Professionals.
\end{acknowledgments}

\end{document}